\documentclass[sigconf, screen, nonacm]{acmart}

\usepackage{array}
\setcopyright{none}
\begin{document}

\title{SURMA: Efficient Fine-Grained Communication Offload for SuperPOD}

\author{Haoran Shang}
\authornote{Haoran Shang and Weifeng Zhang contributed equally to this work and are co-first authors.}
\email{shanghr23@mails.tsinghua.edu.cn}
\affiliation{%
  \institution{Tsinghua University}
  \city{Beijing}
  \country{China}
}

\author{Weifeng Zhang}
\authornotemark[1]
\email{wikkizhang@tencent.com}
\affiliation{%
  \institution{Tencent}
  \country{China}
}

\author{Mengzhu Li}
\email{mz-li24@mails.tsinghua.edu.cn}
\affiliation{%
  \institution{Tsinghua University}
  \city{Beijing}
  \country{China}
}

\author{Huizheng Wang}
\email{wanghz22@mails.tsinghua.edu.cn}
\affiliation{%
  \institution{Tsinghua University}
  \city{Beijing}
  \country{China}
}

\author{Sihan Guan}
\email{guan-sh23@mails.tsinghua.edu.cn}
\affiliation{%
  \institution{Tsinghua University}
  \city{Beijing}
  \country{China}
}

\author{Xinyang Huang}
\email{xhuangci@connect.ust.hk}
\affiliation{%
  \institution{Hong Kong University of Science and Technology}
  \city{Hong Kong}
  \country{China}
}

\author{Bowen Liu}
\email{liubw@ust.hk}
\affiliation{%
  \institution{Hong Kong University of Science and Technology}
  \city{Hong Kong}
  \country{China}
}

\author{Ran Shu}
\email{ran.shu@microsoft.com}
\affiliation{%
  \institution{Microsoft Research}
  \city{Beijing}
  \country{China}
}

\author{Xinru Tang}
\email{tangxr23@mails.tsinghua.edu.cn}
\affiliation{%
  \institution{Tsinghua University}
  \city{Beijing}
  \country{China}
}

\author{Ruoyu Wu}
\email{royywu@tencent.com}
\affiliation{%
  \institution{Tencent}
  \country{China}
}

\author{Yinben Xia}
\email{forestxia@tencent.com}
\affiliation{%
  \institution{Tencent}
  \country{China}
}

\author{Xiang Li}
\email{harveyli@tencent.com}
\affiliation{%
  \institution{Tencent}
  \country{China}
}

\author{Zekun He}
\email{kinghe@tencent.com}
\affiliation{%
  \institution{Tencent}
  \country{China}
}

\author{Yachen Wang}
\email{yachenwang@tencent.com}
\affiliation{%
  \institution{Tencent}
  \country{China}
}

\author{Xianneng Zou}
\email{sandyshuang@tencent.com}
\affiliation{%
  \institution{Tencent}
  \country{China}
}

\author{Jinyi Deng}
\email{dengjinyi@mail.tsinghua.edu.cn}
\affiliation{%
  \institution{Tsinghua University}
  \city{Beijing}
  \country{China}
}

\author{Yang Hu}
\email{hu_yang@tsinghua.edu.cn}
\affiliation{%
  \institution{Tsinghua University}
  \city{Beijing}
  \country{China}
}

\renewcommand{\shortauthors}{Shang et al.}

\makeatletter
\newcommand{\surma@institutionmark}[1]{%
  \ifstrequal{#1}{Tsinghua University}{\dagger}{%
    \ifstrequal{#1}{Tencent}{\ddagger}{\S}}%
}
\let\surma@defaultmkauthors\@mkauthors@iii
\def\@mkauthors@iii{%
  \if@ACM@anonymous
    \surma@defaultmkauthors
  \else
    \gdef\@currentauthors{}%
    \gdef\@currentaffiliation{}%
    \def\@author##1{\def\surma@currentauthor{##1}\let\surma@equalmark\@empty}%
    \def\@authornotemark{\def\surma@equalmark{,*}}%
    \def\@@authornotemark##1{\def\surma@equalmark{,*}}%
    \def\email##1##2{}%
    \def\and{,\space}%
    \def\city##1{\ignorespaces}%
    \def\state##1{\ignorespaces}%
    \def\country##1{\ignorespaces}%
    \def\affiliation##1##2{%
      \def\institution####1{%
        \mbox{\surma@currentauthor\textsuperscript{%
          \ensuremath{\surma@institutionmark{####1}}\surma@equalmark}}\ignorespaces}%
      \ignorespaces##2}%
    \global\setbox\mktitle@bx=\vbox{%
      \noindent\unvbox\mktitle@bx\par\medskip
      \centering\normalfont\large
      \addresses\par\smallskip
      \normalsize
      \textsuperscript{\ensuremath{\dagger}}Tsinghua University, Beijing, China\par
      \textsuperscript{\ensuremath{\ddagger}}Tencent, China\par
      \textsuperscript{\ensuremath{\S}}Hong Kong University of Science and Technology, Hong Kong, China\par
      \smallskip\small
      \textsuperscript{*}Haoran Shang and Weifeng Zhang contributed equally to this work and are co-first authors.\par
      \bigskip}%
  \fi
}
\AtBeginMaketitle{%
  \if@ACM@anonymous\else\let\@authornotes\@empty\fi
}
\makeatother

\begin{abstract}
GPU SuperPODs have become the main platform for large language model (LLM) serving, and as they grow, inter-GPU communication must sustain ever higher bandwidth over ever longer round-trip times (RTTs). On today's GPUs, remote memory access is executed by the SMs themselves, so communication takes two resources from computation: the SMEM or register file (RF) that holds the bandwidth--delay product of data in flight, and the bandwidth of the SM's port to the network-on-chip (NoC), through which every transferred byte passes. Offloading data movement to a dedicated engine frees both, but only if the engine stays efficient for the transfers that LLM serving issues, which are small in total and fine-grained—down to a 576\,B KV-cache entry. NVIDIA's CopyEngine offloads data movement, yet uses less than 1\% of the NVLink bandwidth at a 7\,KB granularity.

We present SURMA, a Scale-Up Remote Memory Access engine on the GPU die. Building it on RDMA faces three challenges: doorbell-based submission is slow; per-application connections need far more state than the die can hold, while on shared connections order-preserving completion notifications block later transactions behind earlier ones; and spraying a transfer over many paths forces either Sender Ordering or Receiver Ordering, which keeps per-transaction state, to keep the completion flag behind the data. SURMA addresses them with a credit-based shared send queue that cuts submission latency from 1.2\,$\mu$s to 400\,ns, fused write-with-atomic semantics that let applications share connections and notify each transaction independently, and Network Per Path Ordering that keeps the flag behind the data with neither Sender Ordering nor Receiver Ordering. We implement SURMA in RTL on a two-FPGA prototype and evaluate it with two 8-GPU NVLink servers. SURMA saturates the attainable bandwidth with a single SM at a 512\,B granularity, whereas NVLink needs over 14\% of the SMs; it occupies under 0.5\% of the area of a GPU and speeds up inference on four models by 1.05--1.26$\times$ over NVLink.
\end{abstract}

\maketitle

\section{Introduction}\label{sec:Introduction}

GPU SuperPODs join tens to thousands of accelerators through a scale-up fabric that exposes the HBM of every device in a global address space, as in NVIDIA NVL72~\cite{nvidia2024gb200nvl72}, NVL576~\cite{nvidia2026vrpod}, and Google Ironwood~\cite{google2025ironwood}. They have become the main platform for LLM training~\cite{shoeybi2019megatron, lepikhin2021gshard}, LLM inference~\cite{deepseekai2025v3, deepep2025, patel2024splitwise, qin2025mooncake}, and scientific computing such as molecular dynamics~\cite{doijade2025gromacs}. While operator-level optimizations reduce computation \cite{wang2026lapa,wang2025mcbp,wang2026bitstopper} and local memory traffic~\cite{wang2025beta,wang2024sofa,wang2026pade,wu2026crosskv}, scaling LLM execution across these systems requires efficient inter-GPU communication. Two trends in SuperPOD evolution make this communication increasingly demanding:
(1) Scale grows, and so does the round-trip time (RTT). An NVL72 connects its GPUs through a single tier of switches, but larger SuperPODs such as NVL576 span racks with a second switch tier~\cite{nvidia2026vrpod}, which adds switch hops and longer links to every round trip.
(2) Bandwidth grows. Per-GPU scale-up bandwidth has risen 12$\times$ in under a decade, from 150\,GB/s per direction with NVLink~2~\cite{nvlink} to 1.8\,TB/s with NVLink~6~\cite{nvidia2026rubin}.

\begin{figure}[!tb]
  \centering
 \includegraphics[width=\linewidth]{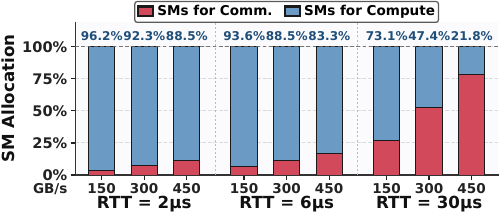}
  \Description{A plot of the modeled fraction of a Hopper GPU's SMs occupied by communication, which rises as RTT and per-direction bandwidth increase.}
  \caption{Modeled share of a Hopper GPU's SMs that communication takes, versus RTT and per-direction bandwidth.}
  \label{fig:sm_alloc}
\end{figure}

\begin{figure}[!tb]
  \centering
 \includegraphics[width=0.92\linewidth]{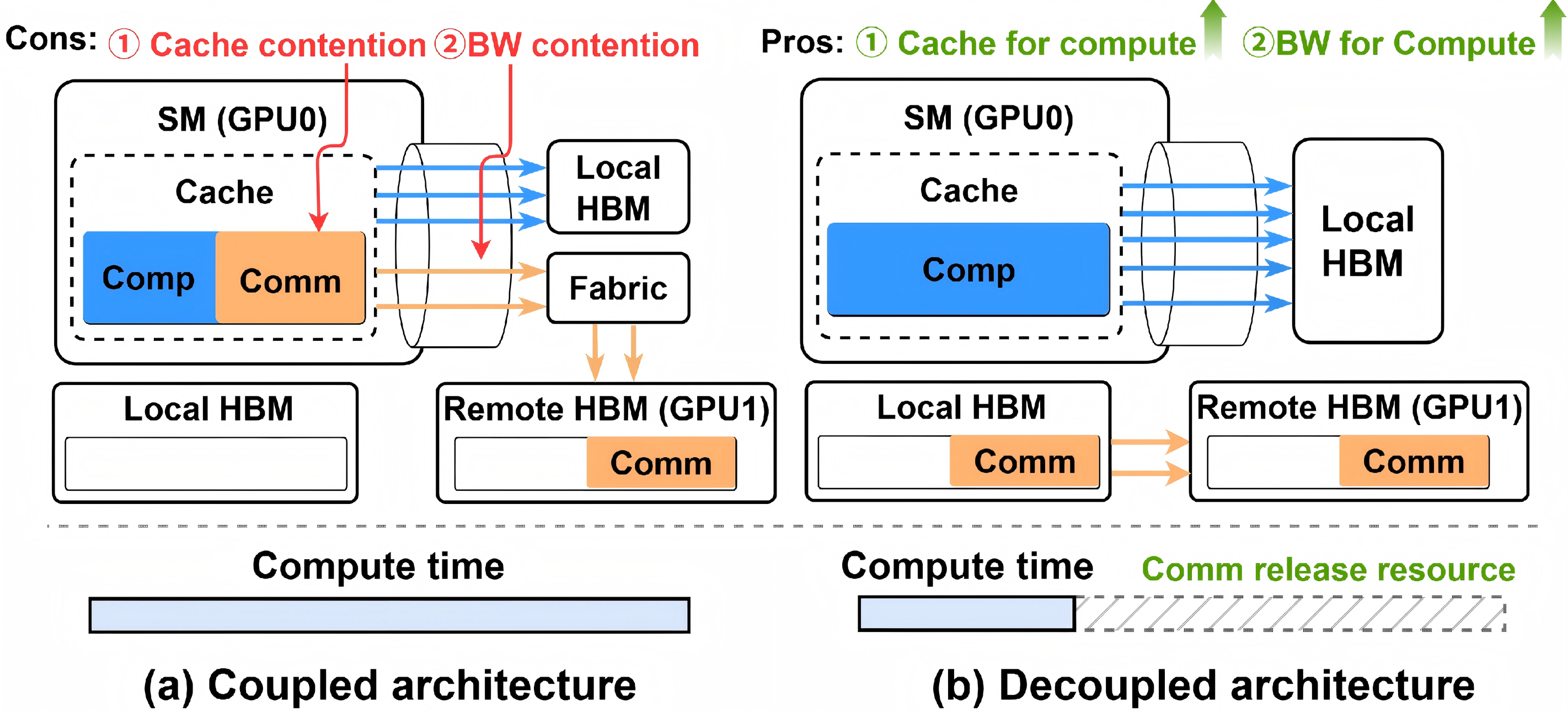}
  \Description{Two block diagrams. In (a), the SMs move data to remote HBM themselves; in (b), the SMs only submit requests and the on-die SURMA engine moves the data from HBM to HBM.}
  \caption{(a) Today's compute--communication-coupled architecture, in which the SMs move the data; (b) SURMA, in which the SMs only submit requests and an on-die engine moves the data from HBM to HBM.}
  \label{fig:base vs surma}
\end{figure}

On current GPUs, both trends land on the SMs. Communication libraries such as NCCL, NVSHMEM, and DeepEP execute remote memory access with the load/store units or the Tensor Memory Accelerator (TMA) of the SMs~\cite{nccl, nvshmem, deepep2025, cudaguide}, so communication takes two SM resources that computation also needs (Figure~\ref{fig:base vs surma}(a)). (1) SMEM or RF. By Little's law, sustaining a bandwidth $B$ over a round trip $RTT$ keeps $B \times RTT$ bytes in flight, the bandwidth--delay product (BDP), and the SM must hold them in its shared memory (SMEM) or register file (RF) until each access completes; we call this the outstanding buffer. (2) The SM's NoC port. Every byte that the SM reads or writes, local or remote, passes through the fixed-bandwidth port that connects it to the network-on-chip (NoC). On an 8-GPU Hopper server, DeepEP needs 12.8\% of the SMs to reach its attainable bandwidth (\S\ref{sec:comm_occupy}), and our model shows this share rising with both RTT and bandwidth (Figure~\ref{fig:sm_alloc}). SM-mediated communication thus couples fabric throughput to compute-side storage and access bandwidth.

Offloading data movement to a dedicated engine decouples the two: the outstanding buffer moves to HBM, and the payload moves from HBM to HBM without passing through the SMs' NoC ports (Figure~\ref{fig:base vs surma}(b)). Such offloading is not new: the CopyEngine (CE) underlies zero-SM collectives in NCCL~2.28~\cite{cudaguide, nccl228}, DMA engines have been proposed to offload ML collectives~\cite{pati2025dmacollectives}, GPU-Initiated Networking (GIN) lets the SMs drive RDMA~\cite{nccl228}, and soNUMA and NIsplit integrate remote memory controllers on CPU dies~\cite{novakovic2014sonuma, daglis2015manycore}. The open question is whether such an engine remains efficient for the transfers of LLM serving, and on a Hopper server existing engines do not: at a 7\,KB granularity, the CE uses less than 1\% of the NVLink bandwidth, and moving 1\,MB takes 1.6$\times$ as long as on the TMA path. RDMA fares no better: even with its bandwidth scaled to match NVLink, a ConnectX-7 RNIC~\cite{connectx7} takes 1.94$\times$ as long as the TMA path to move the same 1\,MB.

These transfers are small and fine-grained. Long contexts limit the batch size, so even with a context of 1M tokens a single transfer moves only a few hundred kilobytes; a token of MoE dispatch carries only 4 or 7\,KB, a KV-cache entry only 576\,B. At this granularity, moving the payload is no longer the expensive part: a 7\,KB token crosses a 450\,GB/s link in about 16\,ns, whereas an RDMA-style engine takes about 1.2\,$\mu$s to accept the request through a doorbell and about 2\,$\mu$s to order its completion flag behind the data. Fine granularity also demands a high request rate: sustaining 450\,GB/s with 576\,B transfers means admitting about $7.8\times10^{8}$ requests per second, or about 50\,GB/s of 64\,B request descriptors. Once a dedicated engine moves the payload, fine-grained efficiency is thus set by how fast the endpoint admits requests, organizes transactions, and publishes completions.

We study this problem in an on-die engine built on RDMA, whose semantics GPUs already use~\cite{deepep2025, nccl228} and whose architecture, unlike the CE's, is open to analysis. Our question is how an on-die remote memory access (RMA) endpoint can share bounded state among many SMs and applications while keeping fine-grained transactions progressing and completing independently. Straightforward designs pay latency or on-die state at three points, quantified for a 512-GPU SuperPOD. (1) \emph{Request admission.} Doorbell submission takes 1.2\,$\mu$s; direct submission into per-QP on-die queues needs 4.5\,GB of SRAM. (2) \emph{Connection sharing.} Per-context connections need 144\,MB of state; sharing them~\cite{tsai2017lite, monga2021flock, zhao2025src} bounds the state, but in-order execution and completion let a 1\,MB transfer delay the small ones behind it by 2.2\,$\mu$s. (3) \emph{Multipath completion.} Keeping the flag behind data sprayed over many paths costs an extra RTT with Sender Ordering or megabytes of per-transaction state with Receiver Ordering, and the GPU's internal bus, unlike PCIe, may still let the flag overtake the data in device memory.

We present SURMA, a Scale-Up Remote Memory Access engine on the GPU die, organized around these three points. (1) \emph{Shared request admission}: the SMs write requests with plain stores into one credit-based shared send queue (SSQ) sized by the BDP of the ports rather than by the number of QPs, and vectorized submission sustains the injection rate. (2) \emph{Non-blocking connection sharing}: all applications share one connection per destination and port, a hardware send sliding window interleaves the resident transactions packet by packet, and write-with-atomic semantics notify each transaction independently. (3) \emph{Path-aware end-to-end completion}: Network Per Path Ordering (NPPO) attaches an atomic to a transaction's tail packet on each in-order path, so the flag follows the data without a source-side wait or per-transaction arrival counters, and destination write ordering tracks only writes still committing to device memory. Two architectural choices underlie these mechanisms: provisioning each class of endpoint state by its own service demand and lifetime, rather than replicating it for every logical communication context, and reusing path-local delivery order to simplify completion.

We implement SURMA in RTL on two Intel Agilex 7 I-Series FPGA boards connected back to back and evaluate it alongside two 8-GPU Hopper servers. The prototype sustains 47.6\,GB/s of its 50\,GB/s line rate at 200\,MHz, or 427\,GB/s when scaled to a 1.8\,GHz GPU clock. With SURMA, a single SM saturates the attainable bandwidth at a 512\,B granularity, whereas SM-driven NVLink transfers need over 14\% of the SMs. Against a doorbell-based RDMA baseline with Sender Ordering, SSQ cuts submission latency from 1.2\,$\mu$s to 400\,ns, and SSQ and NPPO together cut the completion time of transfers of 16\,KB and below by 54\%. Relative to direct submission with per-QP queues and connections, sharing reduces on-die area by 96.2\%, so that SURMA occupies under 0.5\% of a GPU's area for a 512-GPU SuperPOD. In trace-driven evaluation, SURMA speeds up inference on four MoE models by 1.05--1.26$\times$ over NVLink.

\section{Background and Motivation}

\begin{figure}[!tb]
  \centering
 \includegraphics[width=\linewidth]{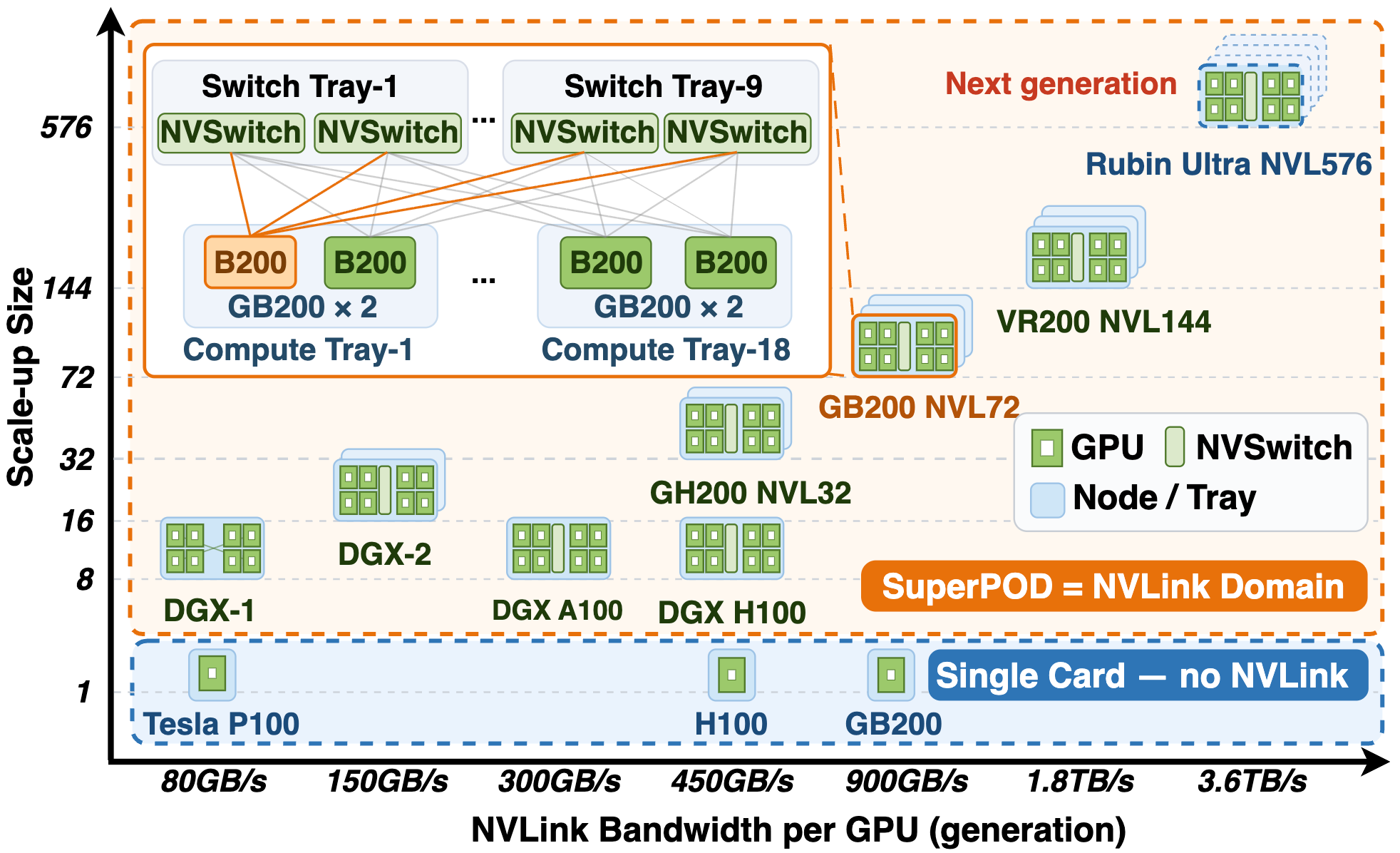}
  \Description{A timeline of NVIDIA SuperPOD generations showing per-GPU NVLink bandwidth and scale-up size growing, with an inset of the one-tier Clos topology of GB200 NVL72.}
  \caption{Evolution of NVIDIA SuperPODs toward higher per-GPU bandwidth and larger scale-up size. The inset shows the one-tier Clos topology of GB200 NVL72.}
  \label{fig:superpod}
\end{figure}

\subsection{GPU SuperPODs}\label{sec:superpod}

LLM deployment calls for SuperPODs. A model whose compute and memory demand exceeds a single GPU~\cite{wang2026temp, wang2026watos} is split by tensor (TP), pipeline (PP), expert (EP), and context parallelism (CP)~\cite{narayanan2021megatron, fedus2022switch, liu2024ringattention}, each of which adds frequent inter-GPU communication—activations for TP and PP, tokens for EP, and KV blocks for CP. A SuperPOD, commonly a cluster interconnected at no less than 200\,GB/s per GPU, answers this demand by tightly coupling many GPUs (Figure~\ref{fig:superpod}). Mixture-of-Experts (MoE) models also demand large SuperPODs: they activate only a few experts per token~\cite{deepseekai2025v3}, so efficient serving spreads the experts over many GPUs with large-scale EP~\cite{zhao2025insights, trtllm2025wideep, wei2026ultraep}, and since EP dispatch and combine span the whole EP group, the group must fit within one SuperPOD.

Each NVIDIA generation raises both bandwidth and scale: from P100 to VR200, per-GPU NVLink bandwidth rose from 80\,GB/s to 1.8\,TB/s, and from DGX-1 to GB200 NVL72 the scale-up size grew from 8 to 72 GPUs~\cite{nvlinkarch, nvidia2026vrnvl72}; Rubin Ultra NVL576 targets 3.6\,TB/s across 576 GPUs. GB200 NVL72 is a one-tier Clos network~\cite{clos1953} of 18 NVSwitch chips, to each of which every GPU connects one NVLink, so any two GPUs are joined by 18 parallel paths of 50\,GB/s each.

\subsection{Communication Occupies the Compute Cores}\label{sec:comm_occupy}

\begin{figure}[!tb]
  \centering
 \includegraphics[width=\linewidth]{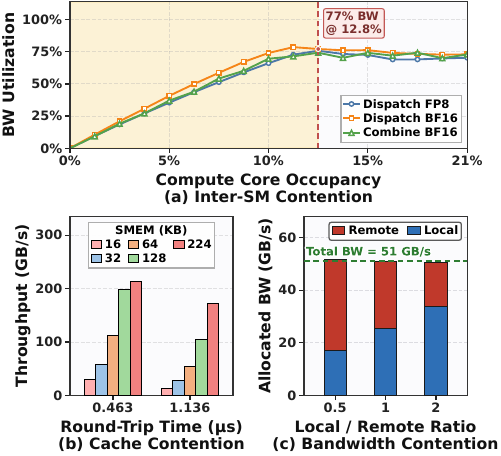}
  \Description{Three panels: (a) DeepEP bandwidth utilization versus the share of SMs allocated to communication; (b) per-SM TMA load throughput versus allocated SMEM for local and remote accesses; (c) the split of one SM's NoC port write bandwidth between local and remote writes.}
  \caption{SM resource contention on a Hopper GPU. (a) DeepEP intranode dispatch (FP8, BF16) and combine (BF16) bandwidth utilization on an 8-GPU server versus communication SM allocation. (b) Per-SM TMA load throughput versus SMEM (16--224\,KB) for local (RTT 0.463\,$\mu$s) and NVLink (RTT 1.136\,$\mu$s) accesses. (c) The SM's 51\,GB/s NoC-port write bandwidth shared by local and remote writes, at local-to-remote ratios of 0.5, 1, and 2.}
  \label{fig:overlap_cost}
\end{figure}

Most existing inter-GPU communication is executed by the compute cores of the GPU: within a SuperPOD, NCCL, NVSHMEM, and DeepEP move data with the load/store units or the TMA of the SMs, so every transfer occupies SMs that computation would otherwise use. We measure this cost with the intranode dispatch and combine kernels of DeepEP on an 8-GPU Hopper server (Figure~\ref{fig:overlap_cost}(a)). The effective bandwidth grows almost in proportion to the SM allocation for communication, and reaching the attainable maximum, about 77\% of the NVLink peak, takes 12.8\% of the SMs.

\begin{figure}[!tb]
  \centering
 \includegraphics[width=\linewidth]{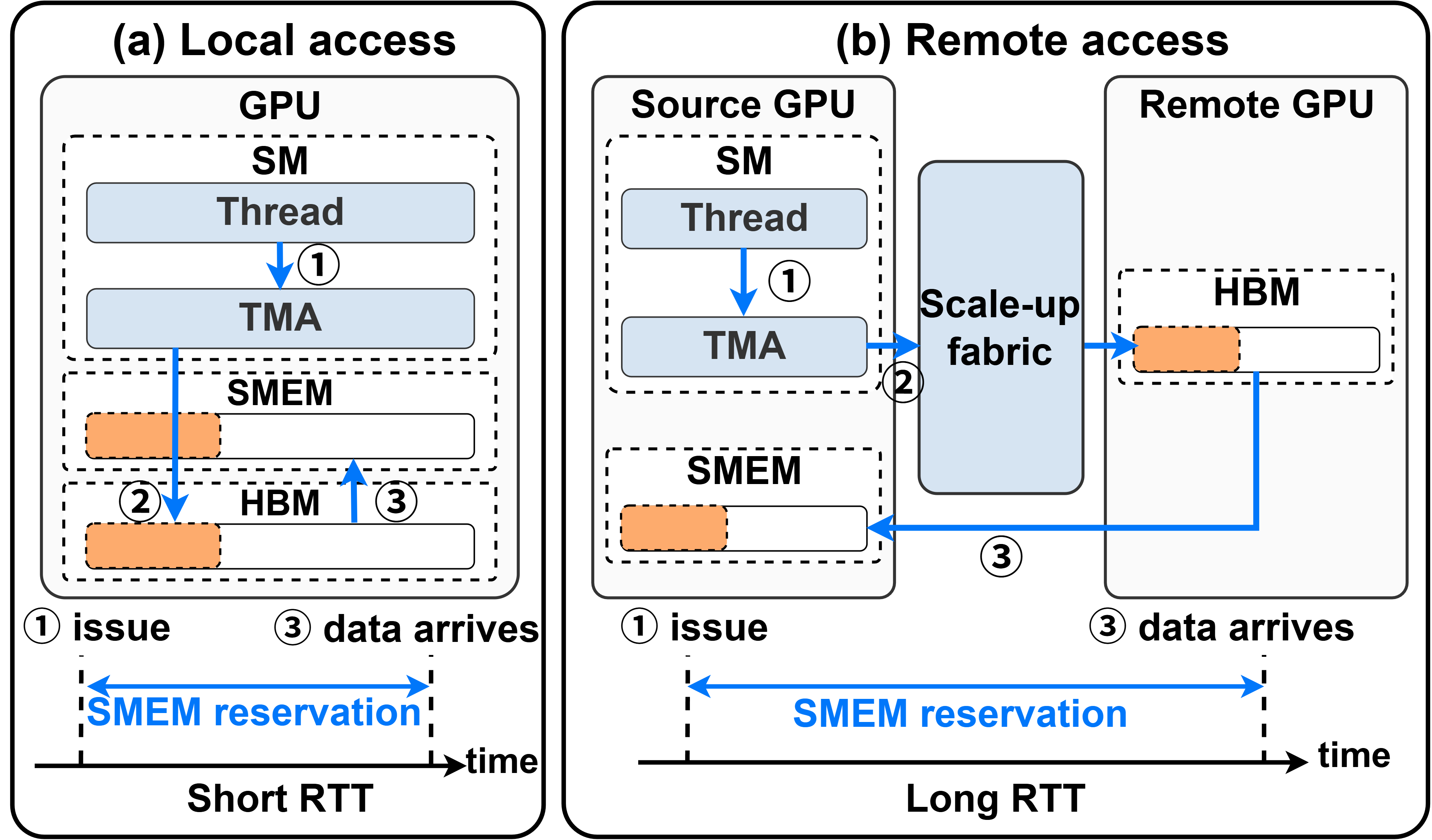}
  \Description{A flow diagram of a TMA load: a thread issues the access to the TMA, the TMA reads from local or remote HBM, and the data returns into the SM's shared memory.}
  \caption{TMA load flow.}
  \label{fig:TMA load}
\end{figure}

We trace this cost by analyzing one TMA load in detail (Figure~\ref{fig:TMA load}): \ding{182} a thread on the SM issues the access to the TMA; \ding{183} the TMA sends a read request to the local HBM or, across the scale-up fabric, to the HBM of a remote GPU; and \ding{184} the data returns and is written into SMEM. Along this path, communication takes two resources that computation also needs. (1) SMEM. The SM must reserve SMEM for the returning data during the whole round trip, so the outstanding buffer of $B \times RTT$ bytes resides in SMEM, and every byte of it is withheld from the tiles of computation. Figure~\ref{fig:overlap_cost}(b) confirms that SMEM caps the throughput: the TMA load throughput of one SM grows almost linearly with the SMEM it is given, and moving from the local RTT of 0.463\,$\mu$s to the NVLink RTT of 1.136\,$\mu$s halves it at 128\,KB, from 197.5 to 104.5\,GB/s. (2) The SM's NoC port. Local and remote accesses traverse the same NoC port of the SM, whose bandwidth is fixed. As Figure~\ref{fig:overlap_cost}(c) shows, local and remote writes divide these 51\,GB/s in proportion to their instruction mix, so every byte that communication moves is subtracted from what computation can access. Based on this analysis, we model how the SM allocation for communication varies with RTT and bandwidth (Figure~\ref{fig:sm_alloc}). When a local hotspot forms in the network, for example at a hot MoE expert, the RTT rises to 30\,$\mu$s and further enlarges the share of the SMs that communication occupies~\cite{pfister1985hotspot, li2019hpcc}.

\subsection{First Attempt: Offloading Communication}\label{sec:offload}

An engine that offloads data movement from the SMs returns both resources of \S\ref{sec:comm_occupy} to computation, but it must transfer small messages efficiently, because the transfers that LLM serving issues have two characteristics. (1) Each transfer moves little data in total. The KV cache of every request occupies a large share of HBM, so long contexts leave room for only a few requests in each batch~\cite{kwon2023vllm}; even with a context of 1M tokens, a single transfer moves only a few hundred kilobytes. (2) The granularity is fine. MoE dispatch and combine move data token by token, and a token carries only 4\,KB in DeepSeek-V4-Pro~\cite{deepseekai2026v4} and 7\,KB in DeepSeek-V3~\cite{deepseekai2025v3}; a KV-cache entry under the multi-head latent attention of DeepSeek-V3 is only 576\,B; and fused kernels that overlap computation with communication send each tile as soon as it is computed~\cite{chang2024flux, zheng2025tilelink, zhang2025comet, sul2026parallelkittens, oh2026perseus}.

Existing offload engines, however, are inefficient for exactly these transfers. The CE keeps the SM cost of communication low, yet our measurements on a Hopper GPU show that at a granularity of 7\,KB it uses less than 1\% of the NVLink bandwidth. RDMA, the other offload path, pays a fixed overhead of 4.64\,$\mu$s per transfer on a Hopper server. As a result, transferring 1\,MB in total takes 3.7\,$\mu$s on the TMA path, but 5.9\,$\mu$s (1.6$\times$) with the CE and, even with the RDMA bandwidth scaled to match NVLink, 7.17\,$\mu$s (1.94$\times$) with a ConnectX-7 RNIC~\cite{connectx7}.

Of the two, RDMA is the more promising base. It is already proven on GPUs, carrying the cross-rack all-to-all of MoE~\cite{deepep2025}, and with GIN in NCCL~2.28, the compute cores of a GPU can drive it directly~\cite{nccl228}. Unlike the CE, RDMA has also been studied extensively, so its architecture is transparent and its bottlenecks, including the fixed overhead above, can be analyzed and addressed. We therefore build SURMA, a remote memory access engine on the GPU die that transfers small messages efficiently within a SuperPOD, on RDMA.

\section{SURMA Overview}

\begin{figure*}[t]
  \centering
 \includegraphics[width=\textwidth]{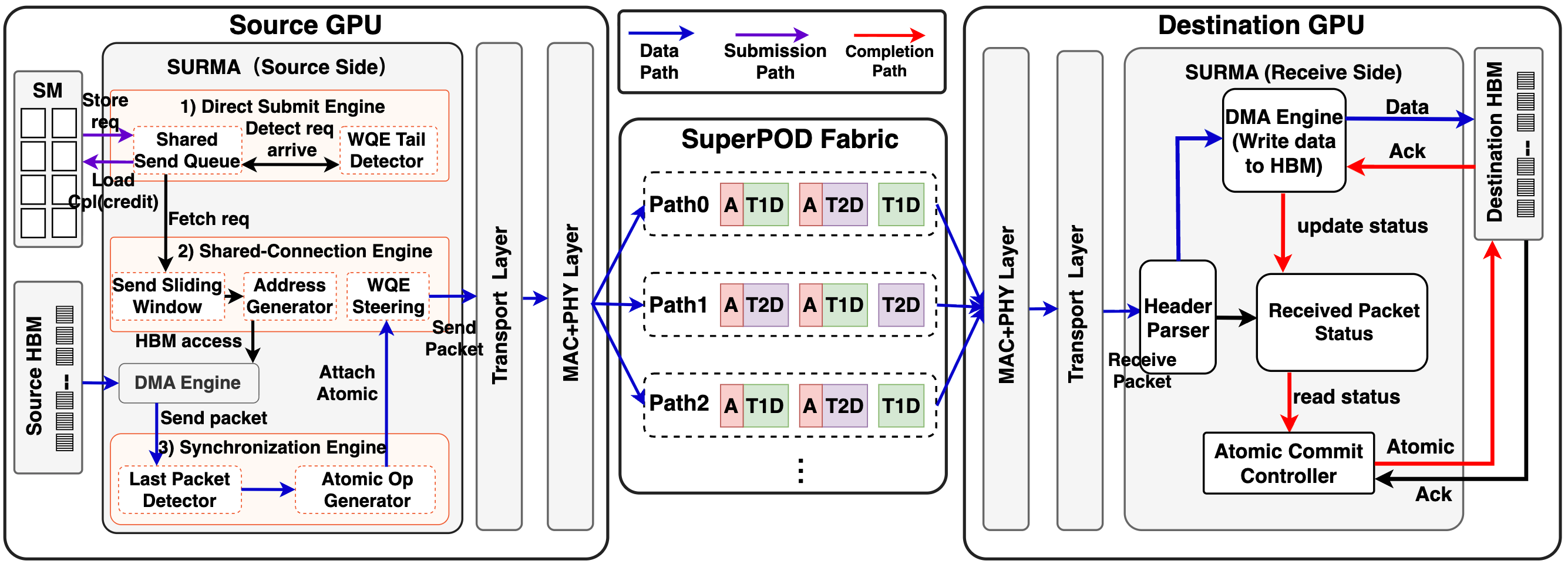}
  \Description{The SURMA hardware architecture at the source and destination GPUs, showing the direct submit engine, the shared-connection engine, the synchronization engine, and the destination components that order writes and atomics.}
  \caption{SURMA hardware architecture. At the source, the direct submit engine admits requests from all SMs (\S\ref{sec:inline doorbell}), the shared-connection engine interleaves transactions over shared connections (\S\ref{sec:SC}), and the synchronization engine attaches atomics to path-tail packets (\S\ref{sec:nppo}). At the destination, the received packet status table and the atomic commit controller hold each atomic until its transaction's writes have committed (\S\ref{sec:nppo}).}
  \label{fig:surma_arch}
\end{figure*}

We present SURMA's workflow and the three challenges of adapting the mature RNIC design to a SuperPOD, then the architecture in Figure~\ref{fig:surma_arch} that addresses them.

\subsection{SURMA Workflow}
\label{sec:workflow}

\begin{figure}[!tb]
  \centering
 \includegraphics[width=\linewidth]{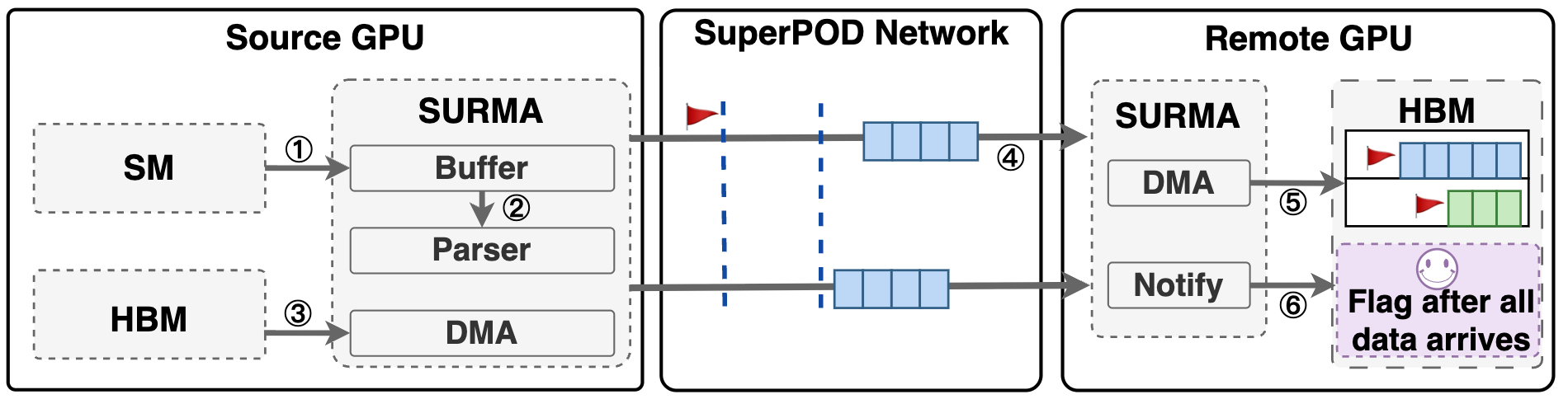}
  \Description{A diagram of the six steps of a SURMA transfer between two GPUs: request submission, request consumption and scheduling, payload transfer, network delivery, remote write, and completion notification.}
  \caption{SURMA workflow.}
  \label{fig:workflow_challenge}
\end{figure}

Each GPU integrates SURMA on its die, with direct access to local HBM and the scale-up fabric, and one transfer proceeds in six steps (Figure~\ref{fig:workflow_challenge}): \ding{182} a source SM posts a compact remote memory access (RMA) request that specifies the operation, source and destination addresses, size, and completion information; \ding{183} the source SURMA parses it and selects fabric paths; \ding{184} it reads the payload from source HBM and injects it into the paths as packets; \ding{185} the fabric delivers the packets to the destination GPU; \ding{186} the destination SURMA writes each payload into destination HBM; and \ding{187} the transaction's completion flag becomes visible only after all of its data has been written, so a consumer that observes it can safely read the data. The latency of one transfer is thus
$$
T_{\mathrm{transfer}} = T_{\mathrm{submit}} + \mathrm{DataSize}/BW + T_{\mathrm{net}} + T_{\mathrm{completion}}.
$$
In an RDMA-based implementation, $T_{\mathrm{submit}} \approx 1.2\,\mu$s and $T_{\mathrm{completion}} \approx 2\,\mu$s exceed both $T_{\mathrm{net}} \approx 1\,\mu$s and, for small messages, $\mathrm{DataSize}/BW$ (\S\ref{sec:three chanllenges}), so SURMA targets these two terms.

\subsection{Design Challenges}
\label{sec:three chanllenges}

From here on, we model a SuperPOD of 512 GPUs, each with 9 scale-up ports of 50\,GB/s per direction—450\,GB/s in total, as on a Hopper GPU—and an RTT of 2\,$\mu$s, typical of a one-tier Clos fabric built on Ethernet; across NVLink we measure 1.136\,$\mu$s (\S\ref{sec:comm_occupy}).

\textbf{Challenge 1: Low-latency, high-rate request admission under bounded on-die state.} The admission path, which carries requests from the SMs into the engine, must meet three requirements at once. (1) \emph{Latency}: a request must enter the engine in far less time than its payload takes to cross the fabric. (2) \emph{Injection rate}: sustaining 450\,GB/s with 576\,B transfers means admitting about $7.8\times10^{8}$ requests per second, and with 64\,B work queue elements (WQEs) the requests alone form a 50\,GB/s stream, comparable to the 51\,GB/s write bandwidth of one SM's NoC port (\S\ref{sec:comm_occupy}). (3) \emph{Bounded buffering}: requests waiting for the engine must fit in on-die SRAM that the GPU can afford.
Existing submission designs trade these requirements against one another. \emph{Doorbell submission}, used by RNICs, keeps the send queues (SQs) in HBM and needs almost no on-die buffer, but the core must write the WQE to HBM, fence, and ring a doorbell over MMIO before the engine fetches the WQE back, which raises $T_{\mathrm{submit}}$ to 1.2\,$\mu$s. Doorbell batching~\cite{dragojevic2014farm, kalia2016design} amortizes the fence and doorbell over many WQEs and raises the rate, but leaves the latency of each request unchanged; BlueFlame~\cite{kalia2016design, rdmacore} carries the WQE in the doorbell write and skips the fetch, but serves only occasional latency-critical messages and falls back to the doorbell once concurrency rises. \emph{Direct submission} lets the SM write each WQE straight into the engine, which cuts the latency but moves the SQs on die. In RDMA, each queue pair (QP) owns one SQ sized to cover one bandwidth--delay product (BDP), typically 256 WQEs or 16\,KB, and each communication thread posts to its own QP to avoid contention, so a GPU needs $N_{\mathrm{QP}} = N_{\mathrm{rank}} \times N_{\mathrm{port}} \times N_{\mathrm{application}} \times N_{\mathrm{thread/application}}$ QPs. With 8 concurrent applications (TP, EP, PP, KV-cache load, \ldots) of 8 threads each, this is $512 \times 9 \times 8 \times 8 \approx 288$\,K QPs, whose SQs take 4.5\,GB.
The SQ capacity thus scales with the number of logical producers, whereas the requests actually outstanding are bounded by the bandwidth and RTT of the ports. A buffer sized by the latter must, however, be written by many SMs at once, and their writes must be kept from colliding without serializing them through a lock or an extra round trip. Request admission must therefore be optimized for both per-request latency and sustained injection rate under a bounded on-die buffer budget.

\textbf{Challenge 2: Connection sharing without transaction serialization.} Each QP is also a reliable connection with its own queue-pair context (QPC) of about 512\,B, such as sequence numbers and acknowledgment and retransmission state. The 288\,K connections of Challenge 1 take 144\,MB of QPC, about 3$\times$ the 50\,MB L2 cache of an H100, and the first two factors of $N_{\mathrm{QP}}$ keep growing with the SuperPOD (\S\ref{sec:superpod}). On the GPU die the connection count must be bounded, and the natural remedy is sharing, as existing RDMA designs do: LITE~\cite{tsai2017lite} shares connections among applications, Flock~\cite{monga2021flock} and SMART~\cite{ren2024smart} share QPs among threads, and SRC~\cite{zhao2025src} multiplexes many QPs onto one connection.
Sharing bounds the state but couples the transactions that share a connection in three ways. (1) \emph{Progress.} A connection serves its transactions in order, so an elephant flow at its head holds back every transaction behind it: a 1\,MB KV-cache transfer occupies the connection for about 2.2\,$\mu$s even at 450\,GB/s, and the few-kilobyte EP transfers queued behind it, each needing under 20\,ns on the wire, wait the full 2.2\,$\mu$s. Flock separates large and small messages onto different QPs~\cite{monga2021flock}, but this adds back the connections that sharing set out to remove. (2) \emph{Completion.} A reliable connection reports completions in posting order, so even after a short transfer's data has arrived, its notification still waits for those of earlier transactions, possibly of other applications. (3) \emph{Flow control.} RDMA runs congestion control per QP; once the transactions of many applications and destinations meet in one queue, backpressure on one connection must not stall transactions bound elsewhere. Any remedy must still let each connection deliver its packets in order, as its transport expects. Reducing connection state is thus useful only when sharing preserves concurrency among independent transactions, in both their progress and their completion.

\textbf{Challenge 3: Low-overhead multipath completion with bounded tracking state.}\label{sec:source block} Under one-sided writes, the receiver learns that a transfer is done only from a completion flag, which must become visible after all of the transfer's data. A SuperPOD joins any two GPUs through many parallel paths—18 in GB200 NVL72 (\S\ref{sec:superpod})—and a transaction sprays its packets across them to use the full bandwidth, so they arrive out of order, and a flag that overtakes its data lets the consumer read stale data. Existing designs order the flag with one of two kinds of evidence. (i) \emph{Sender Ordering} relies on acknowledgments: the source withholds the flag until all of the data has been acknowledged, as ConnectX-7 does~\cite{connectx7}. This adds a full RTT, about 2\,$\mu$s, to every transaction, and the wait blocks later transfers on the same QP, so recovering throughput takes more QPs—exactly the state that Challenge 2 bounds. (ii) \emph{Receiver Ordering} relies on explicit counting: the receiver records, for each transaction, how many packets have arrived against how many it contains, and applies the flag once they match. This removes the extra RTT but costs about 18\,B per in-flight transaction; since in-flight transactions grow with the concurrent applications and the size of the SuperPOD, every receiver must provision several to tens of megabytes for the worst case.
Neither scheme uses what the fabric already guarantees: each path delivers its packets in order, so the last packet of a transaction on a path already attests that all of its earlier packets on that path have arrived. Arrival, however, is not yet visibility. Inside the destination GPU, writes and atomics travel to different L2 slices and HBM channels over different routes, and unlike PCIe, which keeps an atomic behind the writes issued before it, the GPU's internal bus provides no such ordering, so a flag may take effect before data that has already arrived. Completion thus involves two kinds of state with different lifetimes: \emph{network in-flight tracking}, which follows data still crossing the fabric and lives for the transaction's whole flight, and \emph{destination memory-commit tracking}, which follows data that has arrived but is still being written and lives only for the 100--400\,ns of a device-memory write. Path-local ordering can reduce network-level completion tracking, while destination-side memory-commit tracking remains necessary to guarantee application-visible completion.

\subsection{Design Implications}
\label{sec:rationale}

\begin{table}[t]
    \caption{Representative design points under the 512-GPU model, as latency / on-die state. Latency is the submit time, the wait of a small transfer behind a 1\,MB one, and the added completion delay, respectively.}
    \label{tab:tradeoff}
    \centering
    \footnotesize
    \setlength{\tabcolsep}{2pt}
    \renewcommand{\arraystretch}{1.2}

    \begin{tabular}{
        |>{\centering\arraybackslash}m{0.16\columnwidth}
        *{3}{|>{\centering\arraybackslash}m{\dimexpr(0.84\columnwidth-8\tabcolsep-5\arrayrulewidth)/3\relax}}|
    }
        \hline
        \textbf{Phase}
        & \textbf{Low state}
        & \textbf{Low latency}
        & \textbf{SURMA} \\
        \hline

        Submission
        & Doorbell \newline 1.2\,$\mu$s / $\approx$0
        & Per-QP SQ \newline 400\,ns / 4.5\,GB
        & Shared SQ \newline 400\,ns / 144\,KB \\
        \hline

        Connection
        & In-order sharing \newline 2.2\,$\mu$s / 2.25\,MB
        & Per-context \newline 0 / 144\,MB
        & Interleaved sharing \newline $\le$1 pkt/txn / 2.25\,MB \\
        \hline

        Completion
        & Sender ordering \newline 1 RTT / $\approx$0
        & Receiver ordering \newline 0 / MB-scale
        & Path tail + commit \newline 0 / 4.5\,KB \\
        \hline
    \end{tabular}
\end{table}

Table~\ref{tab:tradeoff} recaps the design points analyzed above. In each challenge, a straightforward design either replicates state for every logical communication context or transaction, which the die cannot hold, or gives that state up and lets transactions wait. The table collects representative points rather than a proven frontier: other submission, sharing, and completion schemes exist, and its entries come from different mechanisms under our 512-GPU model. What the challenges share is therefore narrower than a single root cause: the state that a straightforward design provisions grows with logical concurrency, whereas the work that the endpoint serves at any moment is bounded by its ports and the RTT. Two architectural choices in SURMA follow from this observation.

\textit{Provision each class of endpoint state by its own service demand and lifetime.} Rather than replicating every structure for each logical communication context, SURMA sizes each class by what actually bounds it: the shared send queue by the requests outstanding on the ports, about their bandwidth--delay product; connection contexts by destinations and ports; the send window by the transactions scheduled concurrently; and destination commit tracking by the writes still in progress. Not all of this state is fixed by the ports—connection contexts still grow with the number of destination GPUs—but none of it grows with the number of applications or threads. Sharing pays off, however, only if it does not reintroduce the waiting it was meant to avoid: many SMs must write the shared queue without colliding, and transactions on a shared connection must still progress and complete independently (Challenges 1 and 2).

\textit{Reuse path-local delivery order to simplify completion.} Because each path delivers in order, the tail packet of a transaction on a path can stand for the delivery of its earlier packets on that path, so the endpoint need not reconstruct the transaction's full arrival state across paths (Challenge 3). This reduces network-level completion tracking only; the destination still tracks each write until it commits to device memory. A narrower use of the same idea lets the admission path recognize a fully written request by its last store, since the NoC delivers the stores of one SM in order.

\S\ref{sec:design} realizes these choices in shared request admission (\S\ref{sec:inline doorbell}), non-blocking connection sharing (\S\ref{sec:SC}), and path-aware end-to-end completion (\S\ref{sec:nppo}).

\section{Design}
\label{sec:design}

After an overview of how SURMA divides the work of a transaction, we present its three parts: shared request admission (\S\ref{sec:inline doorbell}), non-blocking connection sharing (\S\ref{sec:SC}), and path-aware end-to-end completion (\S\ref{sec:nppo}).

\subsection{Hardware Overview}
\label{sec:overview}

SURMA implements the transaction layer, leaving transport, link, and physical layers to the scale-up interconnect. Figure~\ref{fig:surma_arch} shows how it divides the work of a transaction. The SMs only produce requests: each writes compact WQEs into SURMA and never touches the payload. At the source, SURMA buffers the requests of all SMs in one shared send queue (\S\ref{sec:inline doorbell}), schedules the admitted transactions over shared connections (\S\ref{sec:SC}), reads each packet's payload from local HBM with a DMA engine, and attaches completion atomics to the tail packet on each path (\S\ref{sec:nppo}). At the destination, it writes each payload into device memory and commits each atomic only after its transaction's writes have completed (\S\ref{sec:nppo}). The send queue is shared by all SMs, connection contexts by all applications, and the send window by all resident transactions, whereas each transaction keeps its own identity—a transaction ID, a completion flag, and a record of its outstanding writes—from admission to completion.

\textbf{Data access.} Like the TMA of an SM, SURMA supports 1D, 2D, and 3D transfers, so that the compute cores need not pack non-contiguous data before submitting it. A fused kernel sends each matrix tile as soon as it is computed (\S\ref{sec:offload}), but a tile of a row-major matrix occupies a short segment in each of many rows; with only contiguous transfers, the core would have to pack the tile or submit one WQE per row, again spending its cycles on communication. A WQE instead gives the shape of the transfer once, as the bytes per row and the numbers of rows and planes, together with separate source and destination strides, so a tile can be gathered from one matrix and scattered into another of a different width. \S\ref{sec:implementation} describes the address generator that expands these fields.

\begin{figure}[!tb]
  \centering
 \includegraphics[width=\linewidth]{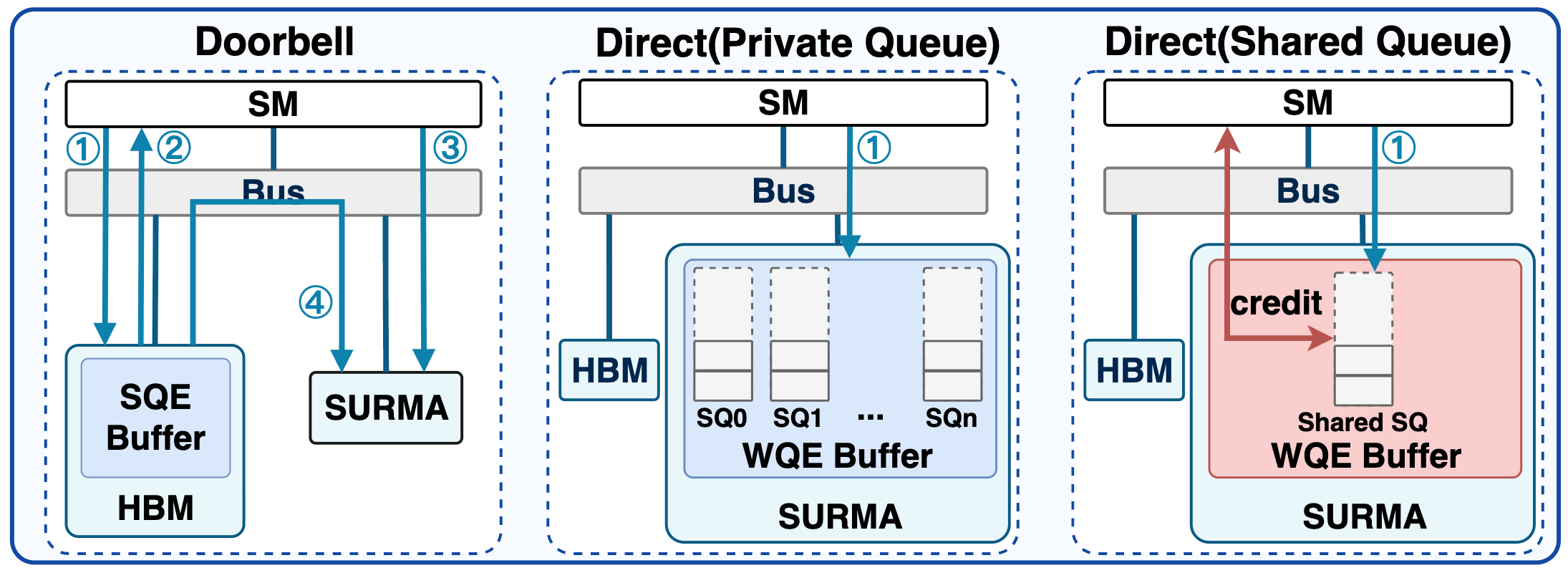}
  \Description{A comparison of request submission strategies: doorbell-based submission, direct submission with per-SM FIFOs, and the credit-based shared send queue shared by all SMs.}
  \caption{Comparisons of different request submission strategies.}
  \label{fig:inline_doorbell}
\end{figure}

\subsection{Shared Request Admission}
\label{sec:inline doorbell}

To minimize submission latency, SURMA adopts direct submission: an SM writes each WQE straight into a WQE buffer inside SURMA that serves as the send queue (SQ) (Figure~\ref{fig:inline_doorbell}), cutting $T_{\mathrm{submit}}$ from 1.2\,$\mu$s to about 400\,ns. Direct submission must then keep the SQs small enough for the die, since one SQ per QP would take 4.5\,GB (\S\ref{sec:three chanllenges}), so SURMA lets many QPs share one SQ, the credit-based shared send queue (SSQ). Sharing raises one basic problem: the tasks of all the QPs on a shared SQ are scheduled together, whereas RDMA runs congestion control independently on each QP, so this unified scheduling must keep the congestion control of one QP from interfering with that of the others. Since the design for Challenge 2 needs a task scheduler anyway, we present this scheduling there (\S\ref{sec:SC}). Beyond this problem, sharing across each factor of $N_{\mathrm{QP}}$ raises its own question, which we address in turn.

\emph{Destinations.} QPs on the same port but to different destinations share one SQ: each WQE carries the ID of its destination GPU, so the engine selects the QP of that destination when it serves the WQE. \emph{Ports.} QPs on different ports share one SQ as well: as with LSU or TMA transfers, applications issue tasks without naming a link, and a single task can use the bandwidth of all ports. Load balancing thus moves into hardware, but since each port leads to an equivalent, independent plane, the hardware simply spreads the load evenly across the ports. \emph{Applications.} Different applications share one SQ once they share connections (\S\ref{sec:SC}). \emph{Threads.} Communication threads run on different SMs, so sharing among threads means letting the SMs share the SQ without contention, which the credit-based flow control below provides. With all four factors shared, one SQ sized by the BDP suffices: 256 outstanding 64\,B WQEs per port, or 144\,KB for 9 ports.

Credit-based flow control lets the SMs share the SQ with plain stores. Once the SMs share one buffer, they contend for its slots: two SMs may write the same slot, or one may overwrite a WQE that has not yet been served. Credit-based flow control avoids such collisions with ordinary store instructions alone, and needs no new instruction. Every address of the WQE buffer is exposed to all SMs, so an SM can write any slot, and software allocates the entries among the SMs on demand—each entry an SM holds is one credit—so that each SM writes only to its own entries and never collides with another. A single SM that performs all of the communication can thus hold every credit and saturate the bandwidth on its own. Each entry carries a Valid bit: hardware sets it to 1 once the WQE has been completely written, and clears it to 0 once the task completes. An SM polls its entries and reclaims an entry, regaining its credit, once the Valid bit returns to 0. Polling needs no dedicated hardware: a single thread polls the Valid bits while the rest of the SM keeps computing, and its cost is that thread's issue slots and the NoC reads of the Valid bits.

Submitting with plain stores leaves one more question: a WQE is larger than a single store, so hardware must tell when an SM has finished writing one. On current GPUs, the NoC that connects the SMs to SURMA uses deterministic routing, so all stores that one SM issues to SURMA arrive in order. An SM therefore writes a WQE with several stores in address order, ending at the entry's last address, and hardware only checks whether an arriving store targets the last address of a WQE: such a store means that the whole WQE has arrived, and hardware sets its Valid bit to 1.

\textbf{Vectorized submission.} Low latency and bounded buffering still leave the injection rate of Challenge 1: sustaining 450\,GB/s with 576\,B transfers takes a 50\,GB/s stream of 64\,B WQEs. Scalar stores may not reach this rate, especially on AI accelerators that devote most of each core's area to vector and matrix units and leave the scalar unit little capability. SURMA therefore also accepts WQEs written with wide vector stores. The core assembles several complete WQEs in its vector registers, where writes stay inside the core, and submits them with a single vector store, which the wide load/store path of the vector unit sustains at a high rate. Software allocates credits for vectorized submission in aligned groups of consecutive entries, so that each vector store fills a fixed number of whole WQEs; since such a store completes all of them at once, hardware recognizes it by its width and sets all of their Valid bits together. Vectorized submission thus raises the request rate without enlarging the queue or adding a submission stage.

\begin{figure}[!tb]
  \centering
 \includegraphics[width=\linewidth]{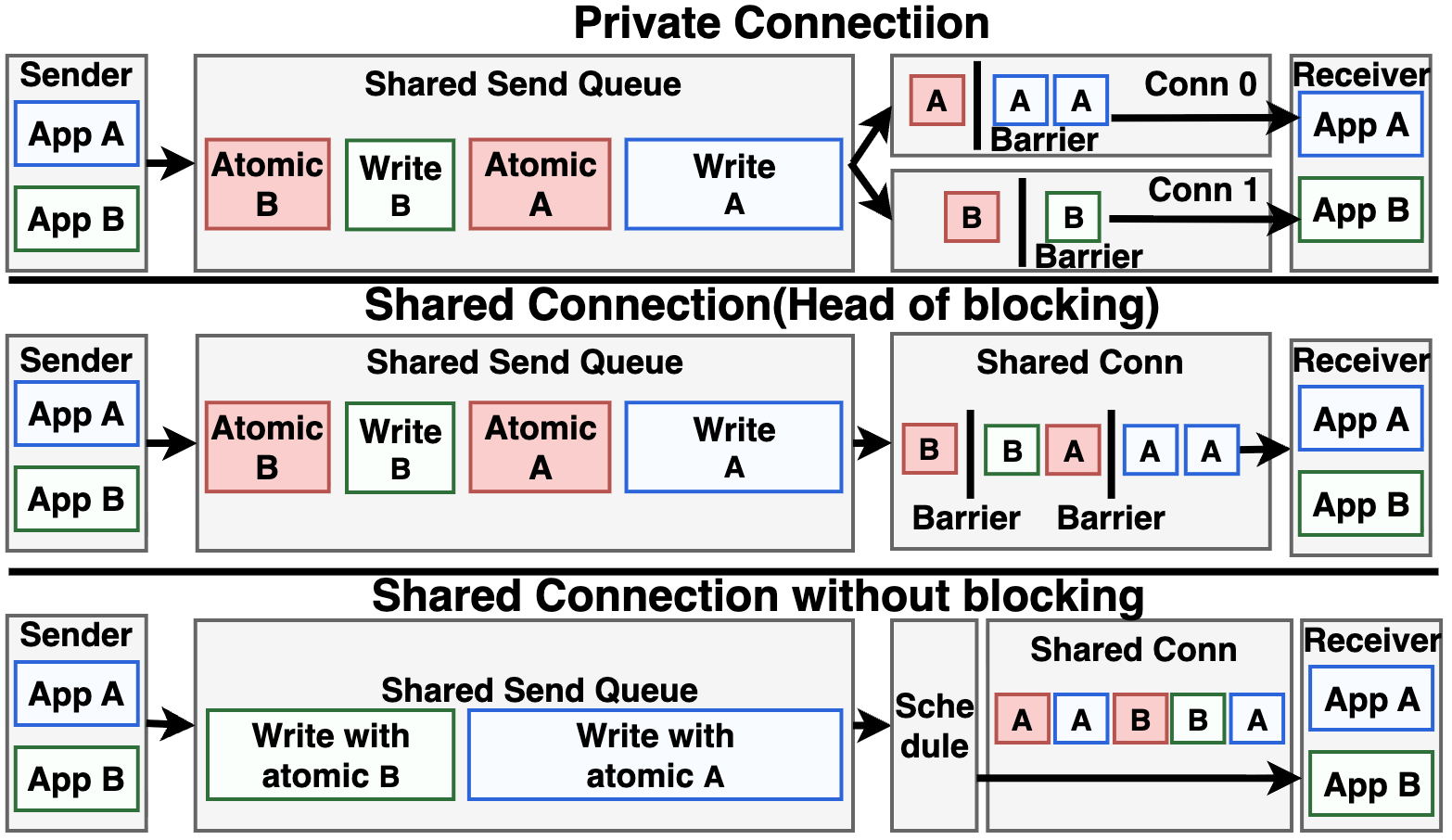}
  \Description{A timeline of transactions on a shared connection: with ordered completion notifications a later transaction is blocked behind earlier ones, whereas with write-with-atomic each transaction is notified as soon as its own data arrives.}
  \caption{Completion notification on a shared connection.}
  \label{fig:shared_connection}
\end{figure}

\subsection{Non-Blocking Connection Sharing}
\label{sec:SC}

To reduce the connection count, SURMA lets all applications—TP, EP, PP, and KV-cache load—share one set of connections, so a GPU needs only one connection per destination and port: 4,608 connections in the SuperPOD of \S\ref{sec:three chanllenges}, whose QPC takes only 2.25\,MB. To keep an elephant flow from blocking the transactions behind it, SURMA splits transactions into packets and interleaves them. Doing so in software would need a central processor, in practice the CPU, that sees the transactions of every SM, adding a GPU--CPU round trip to each transfer and risking a bottleneck, so SURMA schedules in hardware.

SURMA implements a send sliding window in hardware. The window admits transactions in the order in which their requests were written, and serves the resident transactions round-robin: in each turn it selects one transaction and sends one packet of a fixed length, 4\,KB by default to match the MTU of ConnectX-7~\cite{connectx7}. Both the packet length and the window size are configurable. The packet is also the unit that SURMA assigns to ports: successive packets go to the ports in turn. A short transaction thus advances at the same rate as an elephant flow instead of waiting behind it. This interleaving does not break in-order delivery within a connection, because it is scheduling above the connection: once a packet is handed to a connection, the connection still transmits its packets in order and is unaware that they belong to different transactions.

The same window resolves the congestion-control problem of \S\ref{sec:inline doorbell}. SURMA honors per-QP backpressure: when a QP's congestion control backpressures it, the window stops sending packets onto that QP while other QPs continue. It also supports per-destination backpressure. Since all ports lead to equivalent planes, backpressure on one QP likely also affects the other QPs to that destination; the window marks every resident transaction to that destination as ineligible and skips it in the round-robin until the backpressure clears.

\textbf{Independent completion with write-with-atomic.} Interleaving lets the transactions on a shared connection progress independently, but their completions must be independent as well (Challenge 2). With standard RDMA, an application that signals its data posts a write and then an atomic on a flag, and the barrier that keeps the atomic behind its write also holds it behind every earlier operation on the connection, including those of other applications (Figure~\ref{fig:shared_connection}, top). SURMA instead provides a fused write-with-atomic operation: one WQE names both the data and a completion flag in destination memory, and the atomic on that flag is ordered only behind the data of its own transaction. Transactions of different applications therefore complete independently, and the flag of a short transfer can take effect before that of an elephant flow posted earlier on the same connection (Figure~\ref{fig:shared_connection}, bottom). This changes the guarantee that the interface offers: SURMA keeps packets in order within each connection, but unlike a reliable-connection queue pair, it does not order completions across the transactions of a connection. A flag guarantees only that its own transaction's data is visible. If an application needs one transfer to be visible before another, the requester waits until the first transaction completes before submitting the second. Each application polls its own flags, so no completion queue is shared among applications either. Keeping each atomic behind data that is sprayed across paths and written into many device-memory slices is the task of \S\ref{sec:nppo}.

\subsection{Path-Aware End-to-End Completion}
\label{sec:nppo}

\begin{figure}[!tb]
  \centering
 \includegraphics[width=\linewidth]{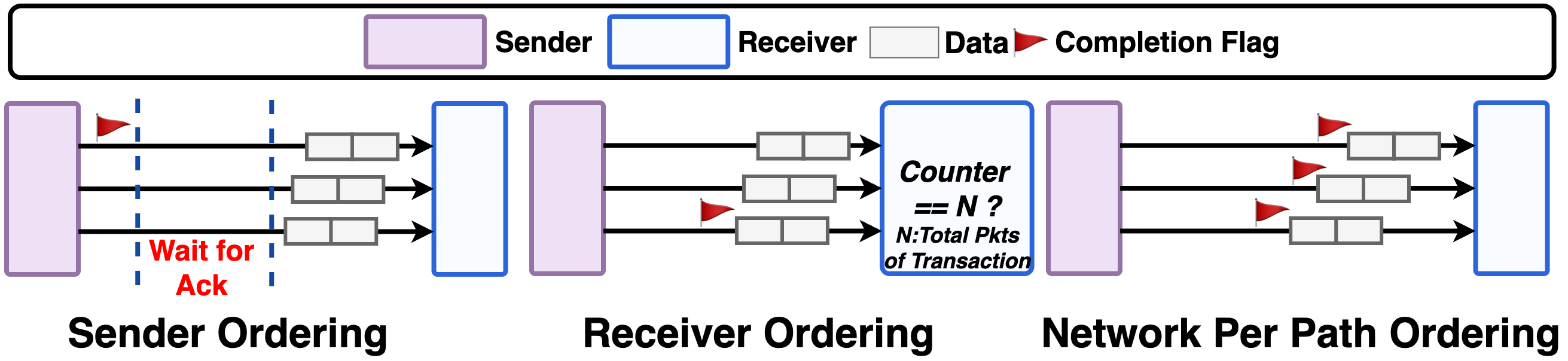}
  \Description{Three schemes for keeping the completion flag of a multipath transaction behind its data: Sender Ordering, Receiver Ordering, and Network Per Path Ordering, which attaches an atomic to the tail packet on each path.}
  \caption{Three ways to order the completion flag behind the data of a multipath transaction.}
  \label{fig:wate}
\end{figure}

\textbf{Network Per Path Ordering (NPPO).} Scale-up protocols such as SUE~\cite{broadcom2025sue} and UALink~\cite{ualink2025} guarantee in-order delivery on a single path, and NPPO derives completion from this guarantee (Figure~\ref{fig:wate}). The source sprays a transaction's packets over the $k$ destination paths in turn, so its last $k$ packets are the tail packets on each path. A Last-Packet Detector marks them, and an Atomic Op Generator attaches an atomic to each. An arriving atomic therefore follows all of the transaction's packets on its path, and once the atomics of all paths have arrived, so has all of the data. With atomic add over 8 paths, each tail packet adds 1, and a flag of 8 signals that all data has been written. Unlike Receiver Ordering (Challenge 3), the receiver keeps no per-transaction count of arrived packets against the transaction's total: an atomic never waits for packets still crossing the fabric.

Some transactions, however, carry too little data to fill every path: a transaction of $n < k$ packets occupies only $n$ paths and would execute fewer than $k$ atomics. The source therefore determines which paths a transaction uses and adjusts the atomic value of its last packet so that the atomics of every transaction sum to $k$: when $n \ge k$, each of the last $k$ packets carries $+1$; when $n < k$, the last packet carries $+(k{+}1{-}n)$ and the others carry $+1$. With 8 paths, a three-packet transaction carries 1, 1, and 6, which still sum to 8. The receiver thus never needs to learn the size of each transaction in advance to know how many atomics to expect, and the application always polls its flag for the same value $k$. Each executed atomic is also acknowledged to the source with its value; the source accumulates the values of a WQE and clears its Valid bit once they reach $k$, so that the SM can reclaim the entry. These acknowledgments only recycle WQE entries: the consumer learns of completion from the atomic itself, so they stay off the critical path.

\textbf{Destination write ordering.} Arrival is not yet visibility: data and atomics reach different slices of device memory over different routes, and the GPU's internal bus does not keep an atomic behind earlier writes (Challenge 3). SURMA therefore records each data packet that it writes into device memory under its transaction, and clears the record once device memory returns the write's completion. Before executing an atomic, SURMA waits until the record shows no outstanding write for that transaction, and then writes the atomic into device memory. Atomics of the same transaction that wait at the same time are merged into a single execution, which adds their combined value in one write. Unlike the counter of Receiver Ordering, this record needs no knowledge of a transaction's size and covers only data that has already arrived and is still being written; such writes typically complete within 100--400\,ns, so only a few transactions are tracked at any time.

\section{Implementation}
\label{sec:implementation}

\textbf{Platform.} We implement SURMA in RTL on two FPGA SmartNIC boards, each built on an Intel Agilex 7 I-Series FPGA with two 200G QSFP56 ports. The boards are connected back to back, so each port carries 25\,GB/s and the two ports together 50\,GB/s in each direction. SURMA runs at 200\,MHz with a 2,048-bit (256\,B) datapath. Since an FPGA hosts neither SMs nor HBM, we emulate both around SURMA. A compute core emulator stands in for the SMs and issues stores to SURMA over an in-order on-chip interconnect, mirroring the deterministic NoC of a GPU (\S\ref{sec:inline doorbell}). A fixed-latency memory model stands in for device memory, and in front of it a memory-side atomic unit executes atomics by read-modify-write, as the atomic units of a GPU's L2 do. SURMA itself implements only the transaction layer. Beneath it, a lightweight custom transport encapsulates each SURMA packet into an Ethernet frame for the 200G Ethernet hard IP and exposes eight logical paths, four on each port, which emulate the multiple paths between two GPUs of a SuperPOD; each path is a FIFO channel and thus delivers in order. Because the boards are directly connected, with no switch to drop packets, and the Ethernet IP corrects bit errors with RS-FEC, the transport omits retransmission; reliability is orthogonal to the transaction layer and is left to the transport of the target fabric.

\textbf{Send path.} Figure~\ref{fig:surma_arch} shows the microarchitecture that the prototype implements. At the source, a direct submit engine, a shared-connection engine, a DMA engine, and a synchronization engine implement \S\ref{sec:inline doorbell}--\S\ref{sec:nppo}. The shared send queue is a 256\,KB buffer of 4,096 64\,B entries; for vectorized submission, software allocates entries in aligned groups of four, which a single 256\,B store fills, and a WQE tail detector sets the Valid bits of one or all four entries on the store that completes them. The send sliding window holds 16 transactions, and each packet carries a 4,032\,B payload and a 64\,B transaction header. For the selected transaction, an Address Generator expands up to three dimensions from the shape and 40-bit strides in the WQE, which still fit in 64\,B, and ends a packet wherever the destination addresses stop being contiguous; the DMA engine then reads the payload from source HBM. On admission, each transaction receives a transaction ID that names it at the receiver. The synchronization engine attaches atomics summing to $k = 8$ to the path-tail packets, and WQE steering prepends the header and hands each packet to the transport layer.

\textbf{Receive path.} At the destination, a header parser passes each payload to a DMA engine and each tail packet's atomic to an atomic commit controller. The received packet status table, which implements the write ordering of \S\ref{sec:nppo}, holds 512 entries of 9\,B, 4.5\,KB in total, keyed by source and transaction ID; the DMA engine increments an entry's counter when it issues a payload write and decrements it when the memory model acknowledges the write. The controller holds each atomic until its counter reaches zero and issues merged atomics as a single atomic add to the memory-side atomic unit. The destination then returns an ACK carrying the atomic's value, and once the values of a WQE reach $k$, the source clears its Valid bit and releases the transaction ID.

\textbf{Resource usage.} Excluding the Ethernet hard IP and the custom transport, SURMA occupies 51,030 ALMs, and its shared send queue takes 256\,KB. Since the two boards form a single pair, the prototype holds connection state only for its eight paths to one destination; the 2.25\,MB of QPC that the 512-GPU configuration of \S\ref{sec:three chanllenges} requires enters the area estimate of \S\ref{sec:eval_area}. SURMA's 2,048-bit datapath at 200\,MHz moves 51.2\,GB/s, enough to sustain the 50\,GB/s of the two ports. Since the throughput of the design scales with its clock, the same RTL clocked at the 1.8\,GHz of a current GPU provides 460.8\,GB/s, enough for the 450\,GB/s NVLink bandwidth of a Hopper GPU, which makes the prototype representative of an on-die deployment.

\section{Evaluation}
\label{sec:evaluation}

\subsection{Testbed}
\label{sec:testbed}

We run our experiments on real GPUs and an FPGA prototype rather than in simulation. On the GPU side we use two 8-GPU NVLink servers built on NVIDIA Hopper GPUs, with 450\,GB/s and 200\,GB/s of NVLink bandwidth and measured local and cross-NVLink latencies of 0.463 and 1.136\,$\mu$s. On the FPGA side we use the two-board prototype of \S\ref{sec:implementation}, which sustains 47.6\,GB/s over 8 paths. End-to-end experiments use four MoE models: Qwen3-Next-80B, Qwen3-235B, GLM4.5-Air, and Ling-flash-2.0.

\subsection{Methodology}

Our evaluation has three parts. (1) How much compute resource communication takes, on the GPUs. Only the throughput of SURMA matters here, so we emulate it with a group of SMs issuing TMA transfers, using HBM as the WQE buffer and the FPGA-measured throughput as their rate. (2) SURMA itself, on the FPGA prototype (\S\ref{sec:implementation}). The compute core emulator on the FPGA models the stores that SMs issue to SURMA with parameters taken from our measurements on a Hopper GPU and from the GPU NoC characterization of Jin et al.~\cite{jin2024gpunoc}. We compare SSQ and NPPO each against the RNIC mechanism it replaces, holding the rest unchanged, so the difference shows whether that mechanism is effective, and we estimate the area that the three mechanisms together add or save. (3) The overall benefit, on both platforms. We collect real-model traces on the GPUs and replace the compute--communication overlap phases with SURMA's measured results.

\subsection{Compute-Resource Occupancy}
\begin{figure}[!tb]
  \centering
 \includegraphics[width=\linewidth]{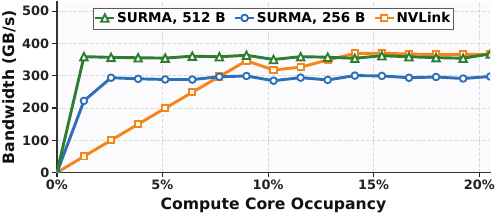}
  \Description{A line plot of peer-to-peer bandwidth versus the fraction of compute cores used for communication. SURMA's curves at 256 B and 512 B flatten after one or two SMs, while NVLink rises roughly linearly.}
  \caption{Peer-to-peer bandwidth versus the fraction of compute cores devoted to communication, for SURMA at transfer granularities of 256\,B and 512\,B and for NVLink transfers by the SMs.}
  \label{fig:surma compute core occupy}
\end{figure}

With SURMA, a single SM saturates the attainable bandwidth at a granularity of 512\,B, and two SMs suffice at 256\,B. Figure~\ref{fig:surma compute core occupy} shows this for a peer-to-peer transfer between two Hopper GPUs, sweeping the fraction of compute cores devoted to communication. SURMA's curves are flat past that point, whereas NVLink climbs roughly linearly and needs well over 14\% of the compute cores to reach the same bandwidth. Its bandwidth is therefore bought with SMs withheld from computation, while SURMA's comes from the on-die engine.

\subsection{Design Effectiveness}

On the FPGA prototype, we evaluate SSQ and NPPO, each against the RNIC mechanism it replaces, and then estimate the area that the three mechanisms of \S\ref{sec:design} together add or save.

\subsubsection{Credit-Based Shared Send Queue}
To isolate the benefit of direct submission, we compare SSQ against a doorbell-based baseline, in which the compute core emulator writes each WQE into memory, posts a fence, and rings a doorbell, and SURMA then fetches the WQE back (\S\ref{sec:three chanllenges}). The baseline's doorbell still travels over the on-chip interconnect rather than PCIe, so the gap we measure comes from direct submission itself and not from a difference in bus performance; everything else is handled exactly as in SURMA.

The 'Base' and 'Base + SSQ' bars of Figure~\ref{fig:ablation} report the comparison, and the gain is concentrated at small transfers: at 16\,KB and below, SSQ cuts the completion time to 0.80. The doorbell interaction costs a fixed amount of time regardless of how much data follows it, so this gain is diluted as the granularity grows, down to 0.99 at 32\,MB. Since real transfers are a few hundred bytes to a few kilobytes (\S\ref{sec:offload}), the mechanism operates precisely where its benefit is largest. Because all SMs share one WQE buffer and coordinate through credits, this gain comes with plain stores and a 256\,KB on-die buffer, whose area we examine next.

\begin{figure*}[t]
  \centering
 \includegraphics[width=\textwidth]{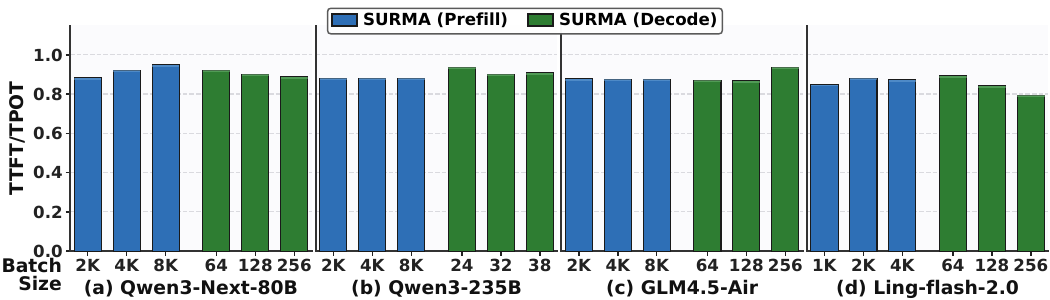}
  \Description{Four groups of plots, one per MoE model, showing SURMA's TTFT for prefill and TPOT for decode normalized to NVLink across batch sizes.}
  \caption{End-to-end inference performance of SURMA, normalized to NVLink (lower is better): TTFT for prefill and TPOT for decode, across batch sizes for four MoE models.}
  \label{fig:e2e}
\end{figure*}

\subsubsection{Area}
\label{sec:eval_area}

\begin{figure}[!tb]
  \centering
 \includegraphics[width=\linewidth]{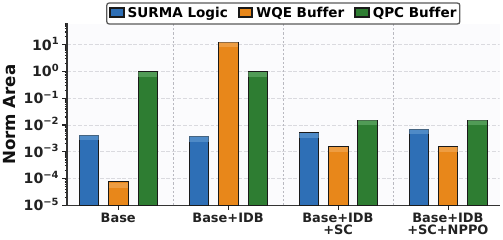}
  \Description{A log-scale bar chart of normalized on-die area as direct submission, shared connections, and Network Per Path Ordering are added to the baseline.}
  \caption{Area breakdown, normalized and in log scale, as direct submission (IDB), shared connections (SC), and Network Per Path Ordering (NPPO) are added to the baseline.}
  \label{fig:area_breakdown}
\end{figure}

\textbf{SURMA reduces the area by 96.2\%.} We measure the resource cost each mechanism adds on the FPGA prototype and use it to estimate the area at TSMC~28\,nm for the 512-GPU configuration of \S\ref{sec:three chanllenges}, reported in Figure~\ref{fig:area_breakdown}. Direct submission alone inflates the WQE buffer sharply, since the send queues move from HBM onto the die. Sharing then brings both the WQE buffer and the QPC buffer down by orders of magnitude, which accounts for the 96.2\% reduction: all SMs share one WQE buffer (\S\ref{sec:inline doorbell}), and interleaved scheduling lets all applications share one set of connections without an elephant flow blocking the transactions behind it (\S\ref{sec:SC}), so the connection count scales only with destinations and ports and the QPC shrinks from about 144\,MB to about 2.25\,MB. NPPO adds only a small amount of logic at the source, and the received packet status table of \S\ref{sec:nppo} stays small. We provision the WQE buffer at 256\,KB rather than the 144\,KB that the bandwidth--delay product asks for (\S\ref{sec:inline doorbell}), so that it absorbs RTT jitter. In total, SURMA occupies under 0.5\% of the area of a GPU.

\subsubsection{Network Per Path Ordering}

\begin{figure}[!tb]
  \centering
 \includegraphics[width=\linewidth]{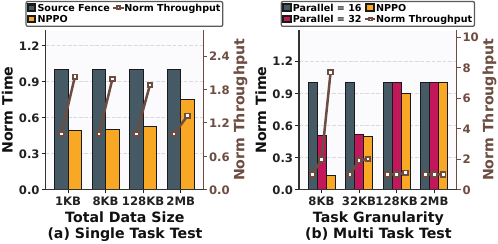}
  \Description{Two panels comparing Sender Ordering with Network Per Path Ordering: (a) normalized completion time of a single transaction across data sizes; (b) normalized throughput with 16 and 32 QPs.}
  \caption{Sender Ordering versus Network Per Path Ordering: normalized completion time and throughput for (a) a single transaction across data sizes and (b) multiple concurrent transactions with 16 and 32 QPs.}
  \label{fig:Inline Doorbell and Write with Atomic}
\end{figure}

We now isolate the completion mechanism, comparing NPPO against the Sender Ordering of existing multipath RNICs (\S\ref{sec:source block}), which withholds the atomic until the acknowledgments for all of the data have returned. Figure~\ref{fig:Inline Doorbell and Write with Atomic}(a) reports the completion time of a single transaction. Below 8\,KB Sender Ordering takes about twice as long, since the round trip it adds is paid once per transaction regardless of how much data follows. That fixed cost is then diluted as the granularity grows, and by 2\,MB the gap narrows to 1.33$\times$.

Figure~\ref{fig:Inline Doorbell and Write with Atomic}(b) turns to sustained throughput, where Sender Ordering has a way to hide this cost: it blocks only the QP it belongs to, so several QPs sharing a path can overlap their waits with one another. The more QPs and the coarser the granularity, the better the concealment—at 8\,KB Sender Ordering is 7.705$\times$ slower with 16 QPs but only 3.878$\times$ with 32, and at 32\,KB doubling the QPs cuts the penalty from 2.009$\times$ to 1.037$\times$, beyond which the two schemes converge. Recovering throughput this way, however, is exactly what inflates the QPC state (\S\ref{sec:three chanllenges}), whereas NPPO needs only one connection per path.

At 2\,MB the speedup falls slightly below one (0.99$\times$), the one point where Sender Ordering is faster: once parallelism has hidden its throughput loss, what remains is the payload bandwidth NPPO spends on the atomic that each path's tail packet carries.

\subsubsection{Mechanisms Breakdown}

\begin{figure}[!tb]
  \centering
 \includegraphics[width=\linewidth]{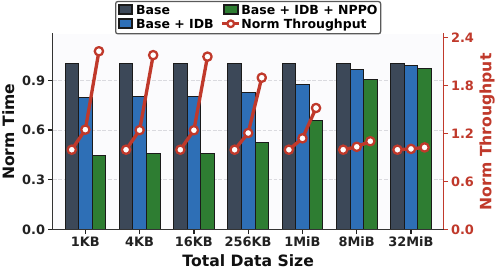}
  \Description{Bar charts of normalized completion time and throughput across data sizes for the RDMA baseline, the baseline with SSQ, and the baseline with both SSQ and NPPO.}
  \caption{Performance breakdown: normalized completion time and throughput across data sizes for the RDMA baseline (doorbell submission with Sender Ordering).}
  \label{fig:ablation}
\end{figure}

The two latency mechanisms together greatly improve the efficiency of small transfers. Adding NPPO on top of SSQ (the 'Base + SSQ + NPPO' bars) brings the completion time of Figure~\ref{fig:ablation} from 0.80 down to 0.46 at 16\,KB and below, so the two together more than double the throughput (2.2$\times$): SSQ shortens $T_{\mathrm{submit}}$ and NPPO shortens $T_{\mathrm{completion}}$, the two terms that dominate small transfers (\S\ref{sec:workflow}). The gain then shrinks as the task grows.

\subsubsection{Comparison with the CopyEngine}
\label{coppyengine}

SURMA outperforms the CE at every transfer size. The CE is the one existing engine that, like SURMA, moves data without occupying any SM. Running a GEMM + AllGather overlap and comparing it against the SM-side TMA path, we find the CE at least 4$\times$ slower whenever the tile of the matrix is 256\,KB or smaller—precisely the regime this paper targets.

\begin{figure}[!tb]
  \centering
 \includegraphics[width=\linewidth]{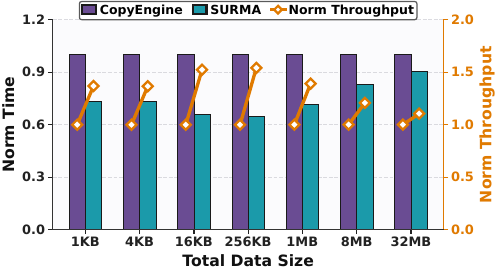}
  \Description{A plot comparing SURMA with the CopyEngine in a GEMM plus AllGather overlap: normalized completion time and SURMA's throughput speedup across data sizes.}
  \caption{SURMA versus the CopyEngine in a GEMM + AllGather overlap.}
  \label{fig:surma_vs_ce}
\end{figure}

Figure~\ref{fig:surma_vs_ce} compares SURMA and the CE on the same workload. SURMA is 1.37--1.55$\times$ faster below 1\,MB, and its advantage narrows to 1.11$\times$ at 32\,MB, where the fixed per-transfer costs no longer matter. At large sizes, the CE's measured peak bandwidth is only 380\,GB/s, 70\,GB/s below the 450\,GB/s NVLink line rate, whereas SURMA reaches 427\,GB/s when its FPGA-measured throughput is scaled to 1.8\,GHz.

\subsection{Overall Performance}

Finally, we evaluate SURMA on overall inference with compute– communication overlap enabled, across four models. We report the speedup of SURMA over NVLink, measured as time-to-first-token (TTFT) for prefill and time-per-output-token (TPOT) for decode, and sweep the batch size in both phases. SURMA improves every configuration, by 1.05–1.26$\times$.

The gains come from two sources, depending on what the phase is bound by. When a phase is compute-bound, the benefit is returning compute resources: communication no longer occupies SMs, so the computation finishes sooner. When it is communication-bound, the benefit is reducing the synchronization overhead: NVLink suffers from the same blocking of Sender Ordering that we analyzed for RNICs in \S\ref{sec:source block}, which NPPO removes.

Figure~\ref{fig:e2e}(b) shows Qwen3-235B. Its prefill is compute-bound and the speedup is essentially constant at about 1.14$\times$: the gain comes entirely from the SMs communication no longer occupies, a quantity independent of the batch size. Decode is instead bound by MoE communication, where the batch size pulls the gain in two directions—a larger batch amortizes each synchronization over more data, shrinking the benefit, but also raises the fraction of time spent communicating, growing it. The latter dominates at first, lifting the speedup from 1.069$\times$ to 1.109$\times$, until the two balance and the curve flattens.

Both phases of Qwen3-Next-80B are communication-bound, but their trends are opposite (Figure~\ref{fig:e2e}(a)). In prefill the speedup declines from 1.131$\times$ to 1.053$\times$, as amortization prevails. In decode it climbs from 1.085$\times$ to 1.126$\times$, because decode's computation is dominated by reading model weights and does not grow with the batch size while its communication volume does, so the communication fraction keeps rising and outweighs the amortization.

\section{Related Work}
\label{sec:related}

\textbf{On-chip Remote Memory Access Engines.} In rack-scale systems, PCIe-attached NICs and their software stacks add microseconds to each remote access, hiding the low latency that fine-grained remote memory access could achieve. soNUMA~\cite{novakovic2014sonuma} therefore integrates the network interface (NI) on the CPU die and exposes one-sided remote accesses; follow-up work scales the NI to manycore chips~\cite{daglis2015manycore}, adds atomic object reads~\cite{daglis2016sabres}, and specializes it for RPCs~\cite{daglis2019rpcvalet, sutherland2020nebula, pourhabibi2021cerebros}. These designs rely on cache-coherent work and completion queues, which GPU SMs lack, and notify a core only after a whole message arrives through Receiver Ordering (\S\ref{sec:source block}), keeping per-message state such as RPCValet's receive counters and NeBuLa's reassembly state; their small-RPC workloads also never face an elephant flow on a shared connection. SURMA follows their on-chip approach but differs in how it admits requests, shares connections, and orders completion (\S\ref{sec:design}).

\textbf{RDMA.} RDMA is the established transport of scale-out networks and can be driven directly by GPUs, which is why SURMA builds on its semantics. Doorbell batching~\cite{dragojevic2014farm, kalia2016design} amortizes the doorbell cost over requests submitted in batches, but does not reduce the latency of an individual request. To contain connection state (Challenge 2), prior work shares connections~\cite{ibta, tsai2017lite, monga2021flock, ren2024smart, zhao2025src, chen2019scalerpc, novakovic2019storm}, uses datagrams~\cite{kalia2016fasst}, or redesigns the NIC~\cite{wang2023srnic, singhvi20201rma}. On a shared connection, however, an elephant flow blocks the transactions behind it; Flock~\cite{monga2021flock} separates large and small messages onto different QPs, which adds back exactly the connections that sharing set out to save.

\section{Conclusion}

SURMA decouples communication from computation in GPU SuperPODs, freeing the SMs' SMEM/register-file capacity and NoC-port bandwidth. With a credit-based shared send queue, shared connections with interleaved scheduling and fused write-with-atomic semantics, and Network Per Path Ordering, this on-die engine saturates attainable bandwidth with a single SM, occupies under 0.5\% of a GPU's area, and speeds up inference by up to 1.26$\times$ over NVLink.

\bibliographystyle{ACM-Reference-Format}
\bibliography{refs}

@inproceedings{wang2024sofa,
  title={SOFA: A compute-memory optimized sparsity accelerator via cross-stage coordinated tiling},
  author={Wang, Huizheng and Fang, Jiahao and Tang, Xinru and Yue, Zhiheng and Li, Jinxi and Qin, Yubin and Guan, Sihan and Yang, Qinze and Wang, Yang and Li, Chao and Yang, Hu and Shouyi, Yin},
  booktitle={2024 57th IEEE/ACM International Symposium on Microarchitecture (MICRO)},
  pages={1247--1263},
  year={2024},
  publisher={IEEE},
  address={Piscataway, NJ, USA}
}

@inproceedings{wang2026temp,
  title={TEMP: A memory efficient physical-aware tensor partition-mapping framework on wafer-scale chips},
  author={Wang, Huizheng and Wei, Taiquan and Wang, Zichuan and Jiang, Dingcheng and Yang, Qize and Liu, Jiaxin and Hou, Jingxiang and Li, Chao and Deng, Jinyi and Hu, Yang and Shouyi, Yin},
  booktitle={2026 IEEE International Symposium on High Performance Computer Architecture (HPCA)},
  pages={1--18},
  year={2026},
  publisher={IEEE},
  address={Piscataway, NJ, USA}
}

@article{wu2026crosskv,
  title={CrossKV: Accelerating Large Language Model Inference via Cross-Stage Dynamic Co-Optimization for KV Cache},
  author={Wu, Nan and Wang, Shenyu and Wang, Huizheng and Wang, Peng and Liu, Xiao and Wang, Zhihua and Hu, Yang and Jiang, Hanjun},
  journal={IEEE Transactions on Circuits and Systems I: Regular Papers},
  year={2026},
  publisher={IEEE}
}

@article{wang2025beta,
  title={BETA: A bit-Grained Transformer attention accelerator With efficient early termination},
  author={Wang, Huizheng and Wang, Hongbin and Yue, Zhiheng and Liu, Jingyao and Wei, Taiquan and Wei, Shaojun and Hu, Yang and Yin, Shouyi},
  journal={IEEE Transactions on Circuits and Systems II: Express Briefs},
  volume={72},
  number={10},
  pages={1433--1437},
  year={2025},
  publisher={IEEE}
}

@inproceedings{wang2026bitstopper,
  title={BitStopper: An efficient Transformer attention accelerator via stage-fusion and early termination},
  author={Wang, Huizheng and Wang, Hongbin and Wei, Shaojun and Hu, Yang and Yin, Shouyi},
  booktitle={2026 31st Asia and South Pacific Design Automation Conference (ASP-DAC)},
  pages={162--169},
  year={2026},
  publisher={IEEE},
  address={Piscataway, NJ, USA}
}

@inproceedings{wang2026lapa,
  title={LAPA: Log-domain prediction-driven dynamic sparsity accelerator for Transformer model},
  author={Wang, Huizheng and Wang, Hongbin and Wei, Shaojun and Hu, Yang and Yin, Shouyi},
  booktitle={2026 31st Asia and South Pacific Design Automation Conference (ASP-DAC)},
  pages={154--161},
  year={2026},
  publisher={IEEE},
  address={Piscataway, NJ, USA}
}

@inproceedings{wang2026pade,
  title={PADE: A predictor-free sparse attention accelerator via unified execution and stage fusion},
  author={Wang, Huizheng and Wang, Hongbin and Wang, Zichuan and Yue, Zhiheng and Wang, Yang and Li, Chao and Hu, Yang and Yin, Shouyi},
  booktitle={2026 IEEE International Symposium on High Performance Computer Architecture (HPCA)},
  pages={1--19},
  year={2026},
  publisher={IEEE},
  address={Piscataway, NJ, USA}
}

@inproceedings{wang2025mcbp,
  title={MCBP: A memory-compute efficient LLM inference accelerator leveraging bit-slice-enabled sparsity and repetitiveness},
  author={Wang, Huizheng and Wang, Zichuan and Yue, Zhiheng and Long, Yousheng and Wei, Taiquan and Yang, Jianxun and Wang, Yang and Li, Chao and Wei, Shaojun and Hu, Yang and Shouyi, Yin},
  booktitle={Proceedings of the 58th IEEE/ACM International Symposium on Microarchitecture},
  pages={1592--1608},
  year={2025},
  publisher={ACM},
  address={New York, NY, USA}
}

@inproceedings{wang2026watos,
  title={WATOS: Efficient LLM training strategies and architecture co-exploration for wafer-scale chip},
  author={Wang, Huizheng and Wang, Zichuan and Wang, Hongbin and Hou, Jingxiang and Wei, Taiquan and Li, Chao and Hu, Yang and Yin, Shouyi},
  booktitle={2026 IEEE International Symposium on High Performance Computer Architecture (HPCA)},
  pages={1--19},
  year={2026},
  publisher={IEEE},
  address={Piscataway, NJ, USA}
}

@article{deepseekai2025v3,
  author  = {{DeepSeek-AI}},
  title   = {{DeepSeek-V3} Technical Report},
  journal = {arXiv preprint arXiv:2412.19437},
  year    = {2025}
}

@online{deepseekai2026v4,
  author       = {{DeepSeek-AI}},
  title        = {{DeepSeek-V4-Pro}},
  year         = {2026},
  url          = {https://huggingface.co/deepseek-ai},
  lastaccessed = {November 8, 2026}
}

@inproceedings{kwon2023vllm,
  author    = {Kwon, Woosuk and Li, Zhuohan and Zhuang, Siyuan and Sheng, Ying
               and Zheng, Lianmin and Yu, Cody Hao and Gonzalez, Joseph E. and
               Zhang, Hao and Stoica, Ion},
  title     = {Efficient Memory Management for Large Language Model Serving
               with {PagedAttention}},
  booktitle = {Proceedings of the 29th Symposium on Operating Systems
               Principles (SOSP)},
  pages     = {611--626},
  year      = {2023},
  publisher = {ACM},
  address   = {New York, NY, USA}
}

@misc{shoeybi2019megatron,
  author        = {Shoeybi, Mohammad and Patwary, Mostofa and Puri, Raul and
                   LeGresley, Patrick and Casper, Jared and Catanzaro, Bryan},
  title         = {{Megatron-LM}: Training Multi-Billion Parameter Language
                   Models Using Model Parallelism},
  year          = {2019},
  archivePrefix = {arXiv},
  eprint        = {1909.08053},
  primaryClass  = {cs.CL},
  doi           = {10.48550/arXiv.1909.08053}
}

@inproceedings{narayanan2021megatron,
  author    = {Narayanan, Deepak and Shoeybi, Mohammad and Casper, Jared and
               LeGresley, Patrick and Patwary, Mostofa and Korthikanti, Vijay
               and Vainbrand, Dmitri and Kashinkunti, Prethvi and Bernauer,
               Julie and Catanzaro, Bryan and Phanishayee, Amar and Zaharia,
               Matei},
  title     = {Efficient Large-Scale Language Model Training on {GPU} Clusters
               Using {Megatron-LM}},
  booktitle = {Proceedings of the International Conference for High
               Performance Computing, Networking, Storage and Analysis (SC)},
  year      = {2021},
  publisher = {ACM},
  address   = {New York, NY, USA}
}

@inproceedings{liu2024ringattention,
  author    = {Liu, Hao and Zaharia, Matei and Abbeel, Pieter},
  title     = {Ring Attention with Blockwise Transformers for Near-Infinite
               Context},
  booktitle = {Proceedings of the 12th International Conference on Learning
               Representations (ICLR)},
  year      = {2024}
}

@inproceedings{lepikhin2021gshard,
  author    = {Lepikhin, Dmitry and Lee, HyoukJoong and Xu, Yuanzhong and
               Chen, Dehao and Firat, Orhan and Huang, Yanping and
               Krikun, Maxim and Shazeer, Noam and Chen, Zhifeng},
  title     = {{GShard}: Scaling Giant Models with Conditional Computation
               and Automatic Sharding},
  booktitle = {Proceedings of the 9th International Conference on Learning
               Representations (ICLR)},
  year      = {2021}
}

@article{fedus2022switch,
  author  = {Fedus, William and Zoph, Barret and Shazeer, Noam},
  title   = {Switch Transformers: Scaling to Trillion Parameter Models with
             Simple and Efficient Sparsity},
  journal = {Journal of Machine Learning Research},
  volume  = {23},
  number  = {120},
  year    = {2022}
}

@inproceedings{patel2024splitwise,
  author    = {Patel, Pratyush and Choukse, Esha and Zhang, Chaojie and
               Shah, Aashaka and G{\"o}ig{\"u}n, In{\i}n{\c{c}} and
               Maleki, Saeed and Bianchini, Ricardo},
  title     = {Splitwise: Efficient Generative {LLM} Inference Using Phase
               Splitting},
  booktitle = {Proceedings of the 51st Annual International Symposium on
               Computer Architecture (ISCA)},
  year      = {2024},
  publisher = {IEEE},
  address   = {Piscataway, NJ, USA}
}

@inproceedings{qin2025mooncake,
  author    = {Qin, Ruoyu and Li, Zheming and He, Weiran and Zhang, Mingxing
               and Wu, Yongwei and Zheng, Weimin and Xu, Xinran},
  title     = {Mooncake: A {KVCache}-centric Disaggregated Architecture for
               {LLM} Serving},
  booktitle = {Proceedings of the 23rd USENIX Conference on File and Storage
               Technologies (FAST)},
  year      = {2025},
  publisher = {USENIX Association},
  address   = {Berkeley, CA, USA}
}

@online{nvlinkarch,
  author       = {{NVIDIA}},
  title        = {{NVIDIA} {Blackwell} Architecture Technical Brief},
  year         = {2024},
  url          = {https://resources.nvidia.com/en-us-blackwell-architecture},
  lastaccessed = {September 9, 2026}
}

@inproceedings{zheng2025tilelink,
  author    = {Zheng, Size and Fang, Jin and Zheng, Xuegui and Hou, Qi and
               Bao, Wenlei and Zheng, Ningxin and Jiang, Ziheng and
               Wang, Dongyang and Ye, Jianxi and Lin, Haibin and
               Chang, Li-Wen and Liu, Xin},
  title     = {{TileLink}: Generating Efficient Compute-Communication
               Overlapping Kernels using Tile-Centric Primitives},
  booktitle = {Proceedings of the 8th Conference on Machine Learning and
               Systems (MLSys)},
  year      = {2025}
}

@inproceedings{zhang2025comet,
  author    = {Zhang, Shulai and Zheng, Ningxin and Lin, Haibin and
               Jiang, Ziheng and Bao, Wenlei and Jiang, Chengquan and
               Hou, Qi and Cui, Weihao and Zheng, Size and Chang, Li-Wen and
               Chen, Quan and Liu, Xin},
  title     = {Comet: Fine-grained Computation-communication Overlapping
               for Mixture-of-Experts},
  booktitle = {Proceedings of the 8th Conference on Machine Learning and
               Systems (MLSys)},
  year      = {2025}
}

@preprint{chang2024flux,
  author        = {Chang, Li-Wen and Bao, Wenlei and Hou, Qi and
                   Jiang, Chengquan and Zheng, Ningxin and Zhong, Yinmin and
                   Zhang, Xuanrun and Song, Zuquan and Jiang, Ziheng and
                   Lin, Haibin and Jin, Xin and Liu, Xin},
  title         = {{FLUX}: Fast Software-based Communication Overlap On {GPUs}
                   Through Kernel Fusion},
  year          = {2024},
  archivePrefix = {arXiv},
  eprint        = {2406.06858},
  primaryClass  = {cs.DC},
  doi           = {10.48550/arXiv.2406.06858}
}

@online{deepep2025,
  author       = {{DeepSeek-AI}},
  title        = {{DeepEP}: An Efficient Expert-Parallel Communication Library},
  year         = {2025},
  url          = {https://github.com/deepseek-ai/DeepEP},
  lastaccessed = {September 9, 2026}
}

@online{nvshmem,
  author       = {{NVIDIA}},
  title        = {{NVSHMEM} Documentation},
  year         = {2025},
  url          = {https://docs.nvidia.com/nvshmem/},
  lastaccessed = {September 9, 2026}
}

@online{nccl,
  author       = {{NVIDIA}},
  title        = {{NVIDIA} Collective Communications Library ({NCCL})},
  year         = {2025},
  url          = {https://developer.nvidia.com/nccl},
  lastaccessed = {September 9, 2026}
}

@online{cudaguide,
  author       = {{NVIDIA}},
  title        = {{CUDA} {C++} Programming Guide},
  year         = {2025},
  url          = {https://docs.nvidia.com/cuda/cuda-c-programming-guide/},
  lastaccessed = {September 9, 2026}
}

@online{nvlink,
  author       = {{NVIDIA}},
  title        = {{NVLink} and {NVLink} Switch},
  year         = {2025},
  url          = {https://www.nvidia.com/en-us/data-center/nvlink/},
  lastaccessed = {September 9, 2026}
}

@online{nvidia2024gb200nvl72,
  author       = {{NVIDIA}},
  title        = {{NVIDIA} {GB200} {NVL72}},
  year         = {2024},
  url          = {https://www.nvidia.com/en-us/data-center/gb200-nvl72/},
  lastaccessed = {October 22, 2026}
}

@online{nvidia2026vrnvl72,
  author       = {{NVIDIA}},
  title        = {{NVIDIA} {Vera} {Rubin} {NVL72}},
  year         = {2026},
  url          = {https://www.nvidia.com/en-us/data-center/vera-rubin-nvl72/},
  lastaccessed = {October 22, 2026}
}

@online{nvidia2026vrpod,
  author       = {{NVIDIA}},
  title        = {{NVIDIA} {Vera} {Rubin} {POD}: Seven Chips, Five Rack-Scale
                  Systems, One {AI} Supercomputer},
  year         = {2026},
  url          = {https://developer.nvidia.com/blog/nvidia-vera-rubin-pod-seven-chips-five-rack-scale-systems-one-ai-supercomputer/},
  lastaccessed = {October 22, 2026}
}

@online{google2025ironwood,
  author       = {{Google}},
  title        = {Ironwood: The First {Google} {TPU} for the Age of Inference},
  year         = {2025},
  url          = {https://blog.google/products/google-cloud/ironwood-tpu-age-of-inference/},
  lastaccessed = {October 22, 2026}
}

@inproceedings{dragojevic2014farm,
  author    = {Dragojevi{\'c}, Aleksandar and Narayanan, Dushyanth and
               Castro, Miguel and Hodson, Orion},
  title     = {{FaRM}: Fast Remote Memory},
  booktitle = {Proceedings of the 11th USENIX Symposium on Networked Systems
               Design and Implementation (NSDI)},
  year      = {2014},
  publisher = {USENIX Association},
  address   = {Berkeley, CA, USA}
}

@inproceedings{kalia2016design,
  author    = {Kalia, Anuj and Kaminsky, Michael and Andersen, David G.},
  title     = {Design Guidelines for High Performance {RDMA} Systems},
  booktitle = {Proceedings of the 2016 USENIX Annual Technical Conference
               (USENIX ATC)},
  year      = {2016},
  publisher = {USENIX Association},
  address   = {Berkeley, CA, USA}
}

@inproceedings{kalia2016fasst,
  author    = {Kalia, Anuj and Kaminsky, Michael and Andersen, David G.},
  title     = {{FaSST}: Fast, Scalable and Simple Distributed Transactions with
               Two-Sided {RDMA} Datagram {RPCs}},
  booktitle = {Proceedings of the 12th USENIX Symposium on Operating Systems
               Design and Implementation (OSDI)},
  year      = {2016},
  publisher = {USENIX Association},
  address   = {Berkeley, CA, USA}
}

@inproceedings{tsai2017lite,
  author    = {Tsai, Shin-Yeh and Zhang, Yiying},
  title     = {{LITE} Kernel {RDMA} Support for Datacenter Applications},
  booktitle = {Proceedings of the 26th ACM Symposium on Operating Systems
               Principles (SOSP)},
  year      = {2017},
  publisher = {ACM},
  address   = {New York, NY, USA}
}

@inproceedings{jin2024gpunoc,
  author    = {Jin, Zhixian and Rocca, Christopher and Kim, Jiho and Kasan, Hans and
               Rhu, Minsoo and Bakhoda, Ali and Aamodt, Tor M. and Kim, John},
  title     = {Uncovering Real {GPU} {NoC} Characteristics: Implications on
               Interconnect Architecture},
  booktitle = {Proceedings of the 57th IEEE/ACM International Symposium on
               Microarchitecture (MICRO)},
  pages     = {885--898},
  year      = {2024},
  publisher = {IEEE}
}

@inproceedings{ren2024smart,
  author    = {Ren, Feng and Zhang, Mingxing and Chen, Kang and Xia, Huaxia and
               Chen, Zuoning and Wu, Yongwei},
  title     = {Scaling Up Memory Disaggregated Applications with {SMART}},
  booktitle = {Proceedings of the 29th ACM International Conference on
               Architectural Support for Programming Languages and Operating
               Systems (ASPLOS)},
  pages     = {351--367},
  year      = {2024},
  publisher = {ACM},
  address   = {New York, NY, USA},
  doi       = {10.1145/3617232.3624857}
}

@inproceedings{zhao2025src,
  author    = {Zhao, Yiren and Shu, Ran and Xiong, Yongqiang},
  title     = {{SRC}: A Scalable Reliable Connection for {RDMA} with Decoupled
               {QPs} and Connections},
  booktitle = {Proceedings of the 9th Asia-Pacific Workshop on Networking
               (APNet)},
  pages     = {44--50},
  year      = {2025},
  publisher = {ACM},
  address   = {New York, NY, USA},
  doi       = {10.1145/3735358.3735366}
}

@inproceedings{chen2019scalerpc,
  author    = {Chen, Youmin and Lu, Youyou and Shu, Jiwu},
  title     = {Scalable {RDMA} {RPC} on Reliable Connection with Efficient
               Resource Sharing},
  booktitle = {Proceedings of the 14th EuroSys Conference (EuroSys)},
  year      = {2019},
  publisher = {ACM},
  address   = {New York, NY, USA}
}

@inproceedings{novakovic2019storm,
  author    = {Novakovic, Stanko and Shan, Yizhou and Kolli, Aasheesh and
               Cui, Michael and Zhang, Yiying and Eran, Haggai and
               Pismenny, Boris and Liss, Liran and Wei, Michael and
               Tsafrir, Dan and Aguilera, Marcos},
  title     = {Storm: A Fast Transactional Dataplane for Remote Data Structures},
  booktitle = {Proceedings of the 12th ACM International Conference on Systems
               and Storage (SYSTOR)},
  year      = {2019},
  publisher = {ACM},
  address   = {New York, NY, USA}
}

@inproceedings{wang2023srnic,
  author    = {Wang, Zilong and Luo, Layong and Ning, Qingsong and
               Zeng, Chaoliang and Li, Wenxue and Wan, Xinchen and Xie, Peng and
               Feng, Tao and Cheng, Ke and Geng, Xiongfei and Wang, Tianhao and
               Ling, Weicheng and Huo, Kejia and An, Pingbo and Ji, Kui and
               Zhang, Shideng and Xu, Bin and Feng, Ruiqing and Ding, Tao and
               Chen, Kai and Guo, Chuanxiong},
  title     = {{SRNIC}: A Scalable Architecture for {RDMA} {NICs}},
  booktitle = {Proceedings of the 20th USENIX Symposium on Networked Systems
               Design and Implementation (NSDI)},
  year      = {2023},
  publisher = {USENIX Association},
  address   = {Berkeley, CA, USA}
}

@inproceedings{singhvi20201rma,
  author    = {Singhvi, Arjun and Akella, Aditya and Gibson, Dan and
               Wenisch, Thomas F. and Wong-Chan, Monica and Clark, Sean and
               Martin, Milo M. K. and McLaren, Moray and Chandra, Prashant and
               Cauble, Rob and Wassel, Hassan M. G. and Montazeri, Behnam and
               Sabato, Simon L. and Scherpelz, Joel and Vahdat, Amin},
  title     = {{1RMA}: Re-envisioning Remote Memory Access for Multi-tenant
               Datacenters},
  booktitle = {Proceedings of the Annual Conference of the ACM Special Interest
               Group on Data Communication (SIGCOMM)},
  year      = {2020},
  publisher = {ACM},
  address   = {New York, NY, USA}
}

@online{ibta,
  author       = {{InfiniBand Trade Association}},
  title        = {{InfiniBand} Architecture Specification, Volume 1, Release 1.7},
  year         = {2023},
  url          = {https://www.infinibandta.org/},
  lastaccessed = {September 9, 2026}
}

@online{connectx7,
  author       = {{NVIDIA}},
  title        = {{NVIDIA} {ConnectX-7} {InfiniBand} Adapter Cards Datasheet},
  year         = {2023},
  url          = {https://www.nvidia.com/en-us/networking/infiniband-adapters/connectx-7/},
  lastaccessed = {September 9, 2026}
}

@online{rdmacore,
  author       = {{Linux RDMA}},
  title        = {rdma-core: {RDMA} Core Userspace Libraries and Daemons
                  ({mlx5} Provider)},
  year         = {2025},
  url          = {https://github.com/linux-rdma/rdma-core},
  lastaccessed = {October 22, 2026}
}

@online{ualink2025,
  author       = {{UALink Consortium}},
  title        = {{UALink} 200G 1.0 Specification},
  year         = {2025},
  url          = {https://ualinkconsortium.org/},
  lastaccessed = {October 22, 2026}
}

@online{broadcom2025sue,
  author       = {{Broadcom}},
  title        = {Scale-Up Ethernet ({SUE}) Framework Specification},
  year         = {2025},
  url          = {https://www.broadcom.com/},
  lastaccessed = {November 6, 2026}
}

@article{clos1953,
  author  = {Clos, Charles},
  title   = {A Study of Non-Blocking Switching Networks},
  journal = {The Bell System Technical Journal},
  volume  = {32},
  number  = {2},
  pages   = {406--424},
  year    = {1953}
}

@inproceedings{novakovic2014sonuma,
  author    = {Novakovi{\'c}, Stanko and Daglis, Alexandros and Bugnion, Edouard and Falsafi, Babak and Grot, Boris},
  title     = {Scale-Out {NUMA}},
  booktitle = {Proceedings of the 19th International Conference on Architectural Support for Programming Languages and Operating Systems (ASPLOS)},
  pages     = {3--18},
  year      = {2014},
  doi       = {10.1145/2541940.2541965},
  publisher = {ACM},
  address   = {New York, NY, USA}
}

@inproceedings{daglis2016sabres,
  author    = {Daglis, Alexandros and Ustiugov, Dmitrii and Novakovi{\'c}, Stanko and Bugnion, Edouard and Falsafi, Babak and Grot, Boris},
  title     = {{SABRes}: Atomic Object Reads for In-Memory Rack-Scale Computing},
  booktitle = {Proceedings of the 49th Annual IEEE/ACM International Symposium on Microarchitecture (MICRO)},
  pages     = {1--13},
  year      = {2016},
  doi       = {10.1109/MICRO.2016.7783709},
  publisher = {IEEE}
}

@inproceedings{daglis2015manycore,
  author    = {Daglis, Alexandros and Novakovi{\'c}, Stanko and Bugnion, Edouard and Falsafi, Babak and Grot, Boris},
  title     = {Manycore Network Interfaces for In-Memory Rack-Scale Computing},
  booktitle = {Proceedings of the 42nd Annual International Symposium on Computer Architecture (ISCA)},
  pages     = {567--579},
  year      = {2015},
  doi       = {10.1145/2749469.2750415},
  publisher = {ACM},
  address   = {New York, NY, USA}
}

@inproceedings{daglis2019rpcvalet,
  author    = {Daglis, Alexandros and Sutherland, Mark and Falsafi, Babak},
  title     = {{RPCValet}: {NI}-Driven Tail-Aware Balancing of $\mu$s-Scale {RPCs}},
  booktitle = {Proceedings of the 24th International Conference on Architectural Support for Programming Languages and Operating Systems (ASPLOS)},
  pages     = {35--48},
  year      = {2019},
  doi       = {10.1145/3297858.3304070},
  publisher = {ACM},
  address   = {New York, NY, USA}
}

@inproceedings{sutherland2020nebula,
  author    = {Sutherland, Mark and Gupta, Siddharth and Falsafi, Babak and Marathe, Virendra and Pnevmatikatos, Dionisios and Daglis, Alexandros},
  title     = {The {NEBULA} {RPC}-Optimized Architecture},
  booktitle = {Proceedings of the 47th Annual International Symposium on Computer Architecture (ISCA)},
  year      = {2020}
}

@inproceedings{pourhabibi2021cerebros,
  author    = {Pourhabibi, Arash and Sutherland, Mark and Daglis, Alexandros and Falsafi, Babak},
  title     = {{Cerebros}: Evading the {RPC} Tax in Datacenters},
  booktitle = {Proceedings of the 54th Annual IEEE/ACM International Symposium on Microarchitecture (MICRO)},
  pages     = {407--420},
  year      = {2021},
  doi       = {10.1145/3466752.3480055},
  publisher = {ACM},
  address   = {New York, NY, USA}
}

@online{nvidia2026rubin,
  author       = {Alvarez, Eduardo and Mehta, Vishal and Ghodsian, Farshad},
  title        = {Inside {NVIDIA} {Rubin} {GPU} Architecture: Powering the Era of Agentic {AI}},
  year         = {2026},
  url          = {https://developer.nvidia.com/blog/inside-nvidia-rubin-gpu-architecture-powering-the-era-of-agentic-ai/},
  note         = {NVIDIA Technical Blog}
}

@article{oh2026perseus,
  author  = {Oh, Byungsoo and Singh, Rachee},
  title   = {Eliminating Hidden Serialization in Multi-Node Megakernel Communication},
  journal = {arXiv preprint arXiv:2605.00686},
  year    = {2026}
}

@inproceedings{sul2026parallelkittens,
  author    = {Sul, Stuart H. and Arora, Simran and Spector, Benjamin F. and R{\'e}, Christopher},
  title     = {{ParallelKittens}: Systematic and Practical Simplification of Multi-{GPU} {AI} Kernels},
  booktitle = {Proceedings of Machine Learning and Systems (MLSys)},
  volume    = {8},
  pages     = {1076--1089},
  year      = {2026}
}

@inproceedings{li2019hpcc,
  author    = {Li, Yuliang and Miao, Rui and Liu, Hongqiang Harry and Zhuang, Yan and Feng, Fei and Tang, Lingbo and Cao, Zheng and Zhang, Ming and Kelly, Frank and Alizadeh, Mohammad and Yu, Minlan},
  title     = {{HPCC}: High Precision Congestion Control},
  booktitle = {Proceedings of the ACM SIGCOMM 2019 Conference},
  pages     = {44--58},
  year      = {2019},
  publisher = {ACM},
  address   = {New York, NY, USA}
}

@article{pfister1985hotspot,
  author  = {Pfister, Gregory F. and Norton, V. Alan},
  title   = {``Hot Spot'' Contention and Combining in Multistage Interconnection Networks},
  journal = {IEEE Transactions on Computers},
  volume  = {C-34},
  number  = {10},
  pages   = {943--948},
  year    = {1985}
}

@inproceedings{doijade2025gromacs,
  author    = {Doijade, Mahesh and Alekseenko, Andrey and Brown, Ania and Gray, Alan and P{\'a}ll, Szil{\'a}rd},
  title     = {Redesigning {GROMACS} Halo Exchange: Improving Strong Scaling with {GPU}-initiated {NVSHMEM}},
  booktitle = {Proceedings of the SC '25 Workshops of the International Conference for High Performance Computing, Networking, Storage and Analysis},
  pages     = {1314--1329},
  year      = {2025},
  doi       = {10.1145/3731599.3767508},
  publisher = {ACM},
  address   = {New York, NY, USA}
}

@online{nccl228,
  author       = {Jeaugey, Sylvain and Bachan, John and Markthub, Pak and He, Zhenhao and Das, Sirshak and Ghodsian, Farshad},
  title        = {Fusing Communication and Compute with New Device {API} and Copy Engine Collectives in {NVIDIA} {NCCL} 2.28},
  year         = {2025},
  url          = {https://developer.nvidia.com/blog/fusing-communication-and-compute-with-new-device-api-and-copy-engine-collectives-in-nvidia-nccl-2-28/},
  note         = {NVIDIA Technical Blog}
}

@inproceedings{monga2021flock,
  author    = {Monga, Sumit Kumar and Kashyap, Sanidhya and Min, Changwoo},
  title     = {Birds of a Feather Flock Together: Scaling {RDMA} {RPCs} with {Flock}},
  booktitle = {Proceedings of the 28th ACM Symposium on Operating Systems Principles (SOSP)},
  year      = {2021}
}

@online{trtllm2025wideep,
  author       = {{NVIDIA TensorRT-LLM Team}},
  title        = {Scaling Expert Parallelism in {TensorRT-LLM} (Part 1: Design and Implementation of Large-scale {EP})},
  year         = {2025},
  url          = {https://nvidia.github.io/TensorRT-LLM/blogs/tech_blog/blog4_Scaling_Expert_Parallelism_in_TensorRT-LLM.html},
  note         = {TensorRT-LLM Technical Blog}
}

@article{wei2026ultraep,
  author  = {Wei, Xinming and Jin, Chao and Dai, Tuo and Zhong, Yinmin and Yu, Shan and Yang, Chengxu and Wu, Bingyang and Zhang, Zili and Mai, Jing and Zhu, Qianchao and Li, Zhouyang and Liu, Yuliang and Luo, Guojie},
  title   = {{UltraEP}: Unleash {MoE} Training and Inference on Rack-Scale Nodes with Near-Optimal Load Balancing},
  journal = {arXiv preprint arXiv:2606.04101},
  year    = {2026}
}

@inproceedings{zhao2025insights,
  author    = {Zhao, Chenggang and Deng, Chengqi and Ruan, Chong and Dai, Damai and Gao, Huazuo and Li, Jiashi and Zhang, Liyue and Huang, Panpan and Zhou, Shangyan and Ma, Shirong and Liang, Wenfeng and He, Ying and Wang, Yuqing and Liu, Yuxuan and Wei, Y. X.},
  title     = {Insights into {DeepSeek-V3}: Scaling Challenges and Reflections on Hardware for {AI} Architectures},
  booktitle = {Proceedings of the 52nd Annual International Symposium on Computer Architecture (ISCA)},
  pages     = {1731--1745},
  year      = {2025},
  doi       = {10.1145/3695053.3731412},
  publisher = {ACM},
  address   = {New York, NY, USA}
}

@article{pati2025dmacollectives,
  author  = {Pati, Suchita and Islam, Mahzabeen and Aga, Shaizeen and Ibrahim, Mohamed Assem},
  title   = {{DMA} Collectives for Efficient {ML} Communication Offloads},
  journal = {arXiv preprint arXiv:2511.06605},
  year    = {2025}
}

\end{document}